\documentclass[amsmath,amssymb,twocolumn, usenatbib,a4paper]{mn2e}   

\usepackage{graphicx}
\usepackage{bm}
\usepackage{multicol}
\usepackage{epsfig}
\usepackage{pifont} 
\usepackage{txfonts}
\usepackage{times}
\usepackage{yfonts}
\usepackage{colortbl}

\fontencoding{T1}

\def\be{\begin{equation}}
\def\ee{\end{equation}}
\def\beq{\begin{equation}}
\def\eeq{\end{equation}}

\def\ba{\begin{eqnarray}}
\def\ea{\end{eqnarray}}

\def\nn{\nonumber}

\newcommand{\LCDM}{$ \Lambda $CDM$\,\,$}
\newcommand{\LCDMnosp}{$ \Lambda $CDM}
\newcommand{\fnl}{f_{\mathrm{NL}}}

\title[Constraining primordial non-Gaussianity with future galaxy surveys]{Constraining primordial non-Gaussianity\\with future galaxy surveys}

\author[T.~Giannantonio,~C.~Porciani,~J.~Carron,~A.~Amara, ~\&~A.~Pillepich]
{Tommaso~Giannantonio$^{1, 2}$\thanks{E-mail: tommaso.giannantonio at universe-cluster.de},
Cristiano~Porciani$^2$, 
 Julien~Carron$^3$,
 Adam~Amara$^3$,
\newauthor
 and  Annalisa~Pillepich$^{4, 3}$\\
$^1$Excellence Cluster Universe, Technical University Munich, Boltzmannstr. 2, D-85748 Garching bei M\"unchen, Germany\\
$^2$Argelander-Institut f\"ur Astronomie der Universit\"at Bonn, Auf dem H\"ugel 71, D-53121 Bonn, Germany\\
$^3$Institute for Astronomy, ETH Zurich, CH-8093 Zurich, Switzerland\\
$^4$UCO/Lick Observatory, Department of Astronomy and Astrophysics, University of California Santa Cruz, Santa Cruz, CA 95064, USA}

\begin{document}

\date{\today}

\pagerange{\pageref{firstpage}--\pageref{lastpage}} \pubyear{2012}

\maketitle

\label{firstpage}

\begin{abstract}
We study the constraining power on primordial non-Gaussianity of future surveys of the large-scale structure of the Universe for both near-term surveys (such as the Dark Energy Survey - DES) as well as longer term projects such as Euclid and WFIRST. Specifically we perform a Fisher matrix analysis forecast for such surveys, using DES-like and Euclid-like configurations as examples, and take account of any expected photometric and spectroscopic data.
We focus on two-point statistics and
we consider three observables: the 3D galaxy power spectrum in redshift space, 
the angular galaxy power spectrum,
and the projected weak-lensing shear power spectrum. 
We study the effects of adding a few extra parameters to the basic \LCDM set. We include the two standard parameters to model the current value for the dark energy equation of state and its time derivative, $w_0, w_a$, and we account for the possibility of primordial non-Gaussianity of the local, equilateral and orthogonal types, of parameter $\fnl$ and, optionally, of spectral index $n_{\fnl}$. 
We present forecasted constraints on these parameters using the different
observational probes. 
We show that accounting for models that include 
primordial non-Gaussianity does not degrade
the constraint on the standard \LCDM set nor on the dark-energy equation of 
state.
By combining the weak lensing data and the information on 
projected galaxy clustering, consistently
including all two-point functions and their covariance,
 we find 
forecasted marginalised errors $\sigma (\fnl) \sim 3$, $\sigma (n_{\fnl}) \sim 0.12 $ from a Euclid-like survey for the local shape of primordial non-Gaussianity, while the orthogonal and equilateral constraints are weakened for the galaxy clustering case, due to the weaker scale-dependence of the bias. In the lensing case, the constraints remain instead similar in all configurations.
\end{abstract}



\section{Introduction}

The accurate measurement of the cosmic microwave background (CMB) anisotropies by the WMAP satellite \citep{Larson:2010gs}, together with the observation of Type Ia Supernovae and the deep and wide mapping of the large-scale structure (LSS) of the Universe, as done by the SDSS survey \citep{Aihara:2011sj}, have led to the standard model of cosmology, whose parameters are now measured with unprecedented accuracy. 
However, the simplest model that fits all observations, \LCDMnosp, has some shortcomings and poorly understood phases, in particular the late-time acceleration (or dark energy) \citep{Frieman:2008sn} and the early-time inflationary phase \citep{Linde:2007fr}.
For these reasons, it was agreed that new generations of surveys should be built to improve upon the current data.  The group known as Dark Energy Task Force (DETF) \citep{2006astro.ph..9591A} has then labelled current finished and ongoing surveys as stages I and II, while future experiments belong to stages III or IV depending on their timescale and forecasted power, such as the Dark Energy Survey (DES) \citep{Abbott:2005bi} and the Euclid satellite \citep{2009arXiv0912.0914L,RedBook} respectively.

These surveys will also shed new light on other basic questions, such as
the origin of the large-scale structure and the
physics of the early universe. 
It is currently assumed that the observed inhomogeneities in the matter density were seeded by quantum fluctuations at primordial times. These fluctuations were then stretched by inflation (or some alternative theory) on super-horizon scales, and acted as seeds first for the observed CMB anisotropies, and then for the matter density fluctuations.
The simplest possible model for the primordial fluctuations consists of a Gaussian random field with nearly scale-invariant power spectrum \citep{Bardeen:1985tr}. However, many models exist for the inflationary phase \citep {Lyth:2009zz}, many of which would alter the statistics of the fluctuations, in particular introducing a non-Gaussian component \citep{Chen:2010xka}. It is therefore important to put observational constraints on the presence (and, once detected, on the amount) of primordial non-Gaussianity (PNG), in order to distinguish between models of the early universe. This is usually done in terms of the bispectrum amplitude $\fnl$ 
for a set of template shapes for the bispectrum of the perturbations  \citep{Liguori:2010hx}
(see the Appendix~\ref{sec:appbias} for a more precise definition).

The most well-established method to constrain PNG is to measure the 3-point statistics of the CMB temperature anisotropies \citep{Komatsu:2010hc}. 
Current limits from  WMAP7 for the local, equilateral and orthogonal types of PNG are $-10 < \fnl^{\mathrm{loc}} < 74$, $-214 < \fnl^{\mathrm{equ}} < 266$ and  $-410 < \fnl^{\mathrm{ort}} < 6$  at 95\% c. l., respectively \citep{Komatsu:2010fb}. This will be greatly improved by the Planck mission, for which the expected uncertainty is of the order $\sigma(\fnl^{\mathrm{loc}}) \simeq 5 $ \citep{Komatsu:2001rj}, limited by cosmic variance.
A complementary way is to constrain PNG from LSS studies. In this case, one can exploit multiple effects. First, the abundance of the most massive structures
existing at any given cosmic time strongly depends on the level of PNG \citep{Lucchin:1987yv,  Pillepich:2008ka}
although the effect is degenerate with other cosmological parameters \citep{PillPorRei}.
The recent detection of a few high-redshift clusters gives some evidence
in favour of non-Gaussian models
\citep{Hoyle:2010ce}, even though no consensus has been reached
\citep{Mortonson:2010mj,2012JCAP...02..009H} also due to the low number statistics.
The abundance of weak lensing peaks has also been proposed as a method to measure PNG \citep{2011ApJ...728L..13M}.
Moreover, in full analogy with the CMB case, 
any 3-point statistic of the galaxy distribution is affected by PNG \citep{Sefusatti:2007ih, Jeong:2009vd}.
However, the situation is complicated by the overlapping effects of primordial and late-time non-Gaussianity, which is driven by the non-linear growth of structure and galaxy biasing \citep{2009astro2010S.158K}. These effects can be disentangled only by a thorough understanding of the non-linear regime.
Similar considerations apply to modifications of  
the matter power spectrum on small scales induced by PNG 
\citep{Taruya:2008pg, Pillepich:2008ka}.
This is why the recent discovery that PNG also generates a strong 
scale-dependence of galaxy biasing on very large scales
\citep{Dalal:2007cu} has attracted particular interest in the literature.
It has been shown that measurements of PNG based on this single 
feature are currently
competitive with CMB studies \citep{Slosar:2008hx, Xia:2011hj}
and will remain so in the future \citep{Afshordi:2008ru,Carbone:2010sb}.

In this paper we want to investigate further these techniques, applying the Fisher matrix formalism to determine to what accuracy we expect future surveys
 will constrain PNG using two-point statistics only. We will compare the constraints from the large- and small-scale effects of the LSS with those expected by the CMB Planck mission, extending the results by \citet{Joachimi:2009ez,Fedeli:2009mt,Carbone:2010sb,Wang:2010gq,Namikawa:2011yr}. In particular, our study
differs from the previous ones as: the forecasts on PNG by \citet{Fedeli:2009mt} only considered weak-lensing data and did not include variations on any other cosmological parameter but $\fnl$; the results by \citet{Carbone:2010sb} only considered the 3D galaxy power spectrum and only the local type of PNG,
while \citet{Joachimi:2009ez,Wang:2010gq} did not consider PNG.

The plan of the paper is as follows: we will review the effects of primordial non-Gaussianity on the LSS in Section \ref{sec:NG} while,
in Section \ref{sec:nonlin},
we will describe how we calculate the non-linear power spectra (for galaxy
clustering and weak-lensing studies) in terms of the cosmological parameters.
We will then describe the future surveys considered in Section \ref{sec:data}, present our basic forecasts in Section \ref{sec:forec}, and expand on them  
in Section \ref{sec:additional}. We shall finally conclude in Section \ref{sec:concl}.

\section {The effects of primordial non-Gaussianity} \label{sec:NG}
\subsection{Definitions}

The simplest single-field, slow-roll model for inflation gives rise to a nearly Gaussian distribution of the curvature perturbations $\zeta$ or the Bardeen potential $\Phi$ at primordial times corresponding to a redshift $z_*$. 
This changes however in most generalisations: many models, and especially multi-field inflation, produce non-Gaussianities \citep[see e.g. the recent review by][]{Byrnes:2010em}.
There are different possibilities for a departure from a purely Gaussian distribution. The most general expression of a deviation from Gaussianity at quadratic level can be expressed by a non-local relationship between the primordial Bardeen's potential $\Phi$ and a Gaussian auxiliary potential $\varphi$. In real space \citep{Schmidt:2010gw}
\be \label{eq:ngbasic}
\Phi (\mathbf{x}, z_*) =  \varphi (\mathbf{x}, z_*) + \left( \fnl * W * \varphi * \varphi \right) (\mathbf{x}, z_*) \, ,
\ee
where the asterisk denotes convolution and $W(\mathbf{y}, \mathbf{z})$ is a kernel whose form describes the type of non-Gaussianity considered. This is often called ``the CMB definition'', as opposed to ``the LSS definition'', as it is written at early times $z_*$.
Here the $\fnl$ function quantifies the amount of PNG at first order, and it represents the first relevant deviation to measure, i.e. the skewness of the perturbations' probability distribution at a given length scale. Subsequent contributions at higher order, e.g. the kurtosis contribution $g_{\mathrm{NL}}$ are also expected in many theories, but will not be considered in the following.
In the simplest models, $\fnl$ does not depend on scale, and thus it simply 
assumes a constant value in Eq.~(\ref{eq:ngbasic}). In most of the following we will assume this simplification, but we will extend the analysis to the scale-dependent case in Section~\ref{sec:nfnl}.

We can define the power spectrum and bispectrum of the potential in the usual way as
\ba \label{eq:powerspecPhi}
\langle \tilde \Phi (\mathbf{k}) \, \tilde \Phi (\mathbf{k'})   \rangle  &=& (2 \pi)^3 \, \delta_D (\mathbf{k}+ \mathbf{k'}) \, P_{\Phi} (k) \, \nonumber \\
\langle \tilde \Phi (\mathbf{k}) \, \tilde \Phi (\mathbf{k'}) \, \tilde \Phi (\mathbf{k''})   \rangle  &=& (2 \pi)^3 \, \delta_D (\mathbf{k}+ \mathbf{k'} + \mathbf{k''}) \, B_{\Phi} (k,k',k'') \, ,
\ea
where the tilde denotes Fourier transformation. Neglecting subdominant corrections from the 4-point correlator (trispectrum), we can also assume $P_{\Phi} (k)  \simeq P_{\varphi} (k) $ at leading order in $\fnl$.
Then it can be seen that applying the definition of Eq.~(\ref{eq:ngbasic}), the kernel $W$ defines the relationship between power spectrum and bispectrum as
\be
B_{\Phi} (k,k',k'') = 2 \fnl \left[ \tilde W (\mathbf{k}, \mathbf{k'}) \, P_{\Phi} (k) \, P_{\Phi} (k') + 2 \, \mathrm{perms.}    \right] \, ,
\ee
assuming constant $\fnl$. 
In the simplest case of PNG of the local type, $W = 1$ and the bispectrum peaks for squeezed triangles ($k  \ll k' \sim k''$); besides this case, we will consider in the following also the equilateral configuration, for which the bispectrum is maximum for $k  \sim k' \sim k''$,  and the orthogonal type, which was constructed to be orthogonal to the previous two types. These configurations or a mixture thereof can be produced under different inflationary scenarios, see e.g. \citet{Babich:2004gb, Senatore:2009gt} for more details. The bispectra for the three cases are given in the Appendix~\ref{sec:appbias}.

 The analysis of CMB bispectra from the Planck satellite is currently considered to be the most promising tool for distinguishing between these scenarios \citep{Fergusson:2008ra}. In this paper we will consider the three configurations separately, although an analysis of a general linear combination of the three modes is in principle possible. 

The primordial Bardeen potential (equal to the gravitational potential with the opposite sign for sub-horizon modes), is then  related  by the Poisson equation to the total matter density perturbations $\delta$. At linear level
\be
\tilde \delta (\mathbf{k}, z) = \alpha (k, z) \, \tilde \Phi (\mathbf{k}, z_*) \simeq \alpha (k, z) \, \tilde \varphi (\mathbf{k}, z_*) \, ,
\ee
where
\be
\alpha (k, z) = \frac {2 \, c^2 \, k^2 \, T(k) \, D(z)} {3 \Omega_m H_0^2} \frac {g(0)} {g(z_*)} \, ,
\ee
and we have introduced the linear growth function $D(z)$ (normalised so
that $D(0)=1$), the
transfer function $T(k)$, and the potential growth function $g(z) \propto (1+z) \, D(z)$.
We can therefore write the tree-level matter
power spectrum
\be \label{eq:p0}
P_0(k,z) = \alpha^2(k,z) P_{\Phi}(k,z_*) \simeq  \alpha^2(k,z) P_{\varphi} (k,z_*).
\ee

\subsection{LSS and primordial non-Gaussianity} \label{sec:lssandprim}
The large-scale structure of the Universe is commonly described in terms of different tracers. In high-density regions of the underlying dark-matter density contrast $\delta$, we observe the formation of dark-matter haloes, whose distribution is described by the field $\delta_h$.
We expect that galaxy formation occurs within these haloes, thus producing
a galaxy density field $\delta_g$. 
On very large scales and
to first approximation, the linear bias coefficients $b_h$ and $b_g$ 
relate the  different density fields as $\delta_h \simeq b_h \, \delta$ and 
$\delta_g \simeq b_g \, \delta$. 
From the observational side,  the dark-matter distribution $\delta$ can be 
directly observed with weak-lensing studies, while $\delta_g$ can be measured 
by mapping the galaxy distribution.
The most natural 2-point statistic that can be observed is then
the power spectrum of these density fields which can be defined in analogy with 
Eq.~(\ref{eq:powerspecPhi}).
Therefore, the galaxy and matter power spectra approximately satisfy 
 $P_g(k)\simeq b^2_g \,P_m(k)$
and a similar relation can be written in terms of $b_h$ 
(which depends on halo mass and time) for the dark-matter
haloes.

The introduction of primordial non-Gaussianity has multiple observable consequences for the two-point statistics. 
First, a small modification of the matter power spectrum appears on small scales \citep{Taruya:2008pg, Pillepich:2008ka}, due to corrections coming from the linear matter bispectrum, which is non-vanishing in this case.
Secondly,  the biasing law between dark-matter halos and the underlying
mass-density field is altered, becoming strongly scale-dependent in the local and orthogonal cases \citep{Dalal:2007cu,Matarrese:2008nc, Slosar:2008hx,Afshordi:2008ru,Valageas:2009vn,Giannantonio:2009ak,Schmidt:2010gw,2011PhRvD..84f1301D}. This is due to the coupling between long- and short-wavelength modes of the perturbations, affecting the halo power spectrum on large scales. In addition to this effect, there is also a smaller scale-independent modification to the bias coming from the non-Gaussian form of the mass functions \citep{Slosar:2008hx, Giannantonio:2009ak}, affecting the  power spectrum on all scales. The above mentioned numerous independent calculations and comparisons with N-body simulations have shown that the 
bias at fixed halo mass can be written as
\be
\label{biasPNG}
b_{\mathrm{eff}} (k, \fnl) = b (\fnl=0) \, + \delta b(\fnl)+
\Delta b(k, \fnl) \, .
\ee
where $\delta b(\fnl)$ and
$\Delta b(k,\fnl)$ denote the scale-independent and the
scale-dependent corrections, respectively.
If we introduce a weighted variance smoothed by a top-hat filter $F_R(k)$
\be
\sigma_{R,n}^2 \equiv \int \frac {d^3k}{(2 \pi)^3} \, k^n \, P_0(k) \, F_R^2 (k) \, ,
\ee
 then the deviation at any given redshift is for each configuration approximately given by \citep{Schmidt:2010gw,2011PhRvD..84f1301D}
\ba \label{eq:dball}
\Delta b^{\mathrm{loc}}(k, \fnl) &\simeq&  \frac{2 \fnl \, \delta_c \, b_L} {\alpha(k)} \,   \nonumber \\
\Delta b^{\mathrm{ort}}(k, \fnl) &\simeq&  -\frac{6 \fnl \,  \, k} {\alpha(k)} \, \frac{\sigma_{R, -1}^2}{\sigma_{R,0}^2} \, \left[ \delta_c \, b_L + 2 \left( \frac{\sigma_{R,-1}' \, \sigma_{R,0}}{\sigma_{R, -1} \, \sigma_{R, 0}'} - 1 \right)  \right]  \nonumber \\
\Delta b^{\mathrm{equ}}(k, \fnl) &\simeq&  \frac{6 \fnl  \, k^2} {\alpha(k)} \,  \frac{\sigma_{R, -2}^2}{\sigma_{R,0}^2}  \, \left[ \delta_c \, b_L + 2 \left( \frac{\sigma_{R,-2}' \, \sigma_{R,0}}{\sigma_{R, -2} \, \sigma_{R, 0}'} - 1 \right)  \right]  \, ,
\ea
where the prime denotes derivatives w.r.t. $R$.
Here the (linear) Lagrangian halo bias $b_L = b (\fnl=0) \, + \delta b(\fnl)-1$ 
includes the (generally small) scale-independent PNG correction.
In the spirit of the halo model, we will assume that the galaxy linear 
bias coefficient $b_g$ 
is given, on large scales, by a weighted average of the halo bias at different masses.
Note that, on scales larger than $\sim 1$Mpc, which we only consider in our analysis,  the scale dependence is translated unaffected from the halo to the galaxy bias. In detail, for any choice of galaxy number density $n(M)$ and halo occupation distribution (HoD) of mean number $N_g(M)$, we can write:
\be
b_g(k) \propto \int b_h(k,M) \, n(M) \, N_g (M) \, dM \, ,
\ee
which means that, independently from the HoD model, the behaviour of the galaxy bias is the same as for the halo bias:
\be
b_g(k,\fnl) = b_g(\fnl=0) + \delta b_g (\fnl) + \Delta b_g (k, \fnl).
\ee

Therefore, when dealing with galaxy clustering,
we will rewrite the Lagrangian bias as $b_L=b-1$ with $b$
a nuisance parameter to be marginalised over (the index $g$ is understood).
From the expressions above, we can see that the asymptotical scale dependence is $\propto k^{-2}, k^{-1}, k^0$ for the three configurations respectively.

\subsubsection*{Fudge factors}

Finally, a note of caution. The Eqs. (\ref{eq:dball}) have been derived
using the peak-background split technique to the halo mass function obtained
through a Press-Schechter ansatz: i.e.
assuming that linear
density perturbations on a given mass scale collapse into dark-matter haloes
if $\delta>\delta_c$.
To first approximation, it is often assumed that $\delta_c\simeq 1.686$
(independent of halo mass) as
expected from the spherical collapse model in an Einstein-de Sitter universe.
However,
more rigorous applications of the excursion-set approach show that
the correction to the halo mass function for primordial non-Gaussianity
is more complicated than this.
Using models of triaxial collapse, \citet{2009MNRAS.398.2143L} found that
the critical threshold is mass-dependent and tends to $\sqrt{0.7}\delta_c$
for large halo masses (where the factor 0.7 is determined by fitting numerical
simulations). Alternatively, accounting for the fluctuations of the collapse
threshold measured in N-body simulations by \citet{2009ApJ...696..636R}, gives an effective threshold $0.89 \delta_c$
\citep {Maggiore:2009hp,Maggiore:2009rx}.

Inspired by this theoretical work, in numerical applications,
the standard collapse threshold $\delta_c$ is sometimes rescaled by a fudge factor $q$; this is a coefficient of order unity which is introduced heuristically to improve
the agreement of the analytical models for the halo mass function and bias
with N-body simulations
\citep{Carbone:2008iz, Grossi:2009an, Pillepich:2008ka, 2010AdAst2010E..89D,Wagner:2011wx}. However, there are several ambiguities concerning this factor.
First, there is no consensus on the actual role of $q$.
Some authors use it in
the ratio (Eq. \ref{LVratio}
in the Appendix~\ref{sec:appbias}) between the halo mass functions
derived from Gaussian and non-Gaussian initial conditions
\citep{Pillepich:2008ka, Giannantonio:2009ak}. Using the peak-background split
to derive the halo bias, this gives Eq. (9) with the replacement
$\delta_c \to q \delta_c$ \citep{Giannantonio:2009ak}.
On the other hand, other authors use the correction $\sqrt{q} \delta_c$
in the halo mass function and $q \delta_c$ in the halo bias
\citep{Grossi:2009an}. Furthermore,
it is sometimes assumed that there are two independent fudge factors
for the mass functions and the bias
\citep{Wagner:2011wx}.
Secondly, even assuming that $q$ is not mass and redshift dependent,
significantly different values of $q$ (with nearly 30 percent deviations)
are required to reproduce
the N-body results when dark-matter haloes in the simulations
are identified using either the
friends-of-friends (FoF) or the spherical overdensity (SO) method:
this makes unclear which value should be actually used for galaxies \citep{Desjacques:2010jw}.
Finally, the improved model by \citet{2011PhRvD..84f1301D} that we adopt, shows good agreement with N-body simulations without the need for any fudge factor, at least for a specific choice of the halo finder.
Because of these reasons, and because the theoretical motivation favouring a
specific value of $q$ is still lacking, in this analysis we will set $q=1$ throughout. This choice will affect our models for the non-linear mass
power spectrum (through
the halo mass function in the halo model) and for the galaxy power spectrum
(through both the halo mass function and the scale dependence of the
galaxy bias coefficient).
The impact of $q$ on the weak-lensing observables is subtle
and somewhat degenerate with other cosmological parameters.
In fact, in Eq. (\ref{LVratio}) $\fnl$ is multiplied by a polynomial
in $q$ whose coefficients depend on the mass variance.
On the other hand, galaxy clustering studies will constrain the amplitude of
PNG almost entirely from the scale-dependent bias on the largest scales.
This implies that observations will basically
constrain the product $q\fnl$ and, at least in this case,
it will be trivial to correct our results
whenever future progress on the fudge factor $q$ will be made.
Another possibility would be to use $q$ as a nuisance parameter to be
marginalised over after assuming a theoretically based prior. Even though
we will not consider this option in this paper, it is clear that this marginalisation would degrade the constraints on $\fnl$ by a similar amount to the
uncertainty in $q$.

\section {The non-linear regime} \label {sec:nonlin}

In order to take advantage of the extended range of measurements which will be available from future surveys, it is desirable to extend the analyses as far as possible into the non-linear regime. 
For the galaxy clustering, the presence of non-linear biasing complicates the issue, so to be conservative, it is reasonable to expect an accurate theoretical modelling only up to scales $k_{\max} \simeq 0.15 h/$Mpc at $z=0$, even though new theoretical methods are being proposed which may extend the accuracy of the predictions to smaller scales, such as e.g. by \citet{Simpson:2011vn}.
However, in the case of weak lensing, the calculations are made simpler by the absence of the bias, and a reasonably accurate modelling is possible up to the smallest scales, if the effect of baryons can be neglected or accurately modelled
(see e.g. \citet{vanDaalen:2011xb} for a quantitative analysis of this issue).
As in most forecast papers appeared so far we will not consider this effect 
in detail here.

\subsection {Halo model}

We compute the non-linear mass power spectrum using the halo model
developed by \citet{Ma:2000ik,Seljak:2000gq} based
on an original idea by \citet{1977ApJ...217..331M}. See \citet{Cooray:2002dia} for a review.
In this approach all the matter in the Universe is assumed to be concentrated
in a set of discrete and clustered dark-matter halos.
The non-linear  power spectrum can be  written as a sum of two terms:
\be \label{eq:P1P2}
P_{m}(k,z, \fnl) = P_1(k,z, \fnl) + P_2(k,z, \fnl).
\ee
The two-halo term $P_2$ dominates on large scales, and represents the correlations between mass concentrations lying in different haloes, while the one-halo term $P_1$ describes correlations of particles which belong to the same halo. These two terms can be calculated as
\ba
P_1(k, z, \fnl) &=& \int_0^{\infty} n(M,z, \fnl) \left[ \frac{\tilde \rho (M, z, k)}{\rho_m} \right]^2 dM  \nonumber \\
P_2(k, z, \fnl) &=& \left[ \int_0^{\infty} n(M,z, \fnl) \, b(M,z,k, \fnl) \, \frac{\tilde \rho (M, z, k)}{\rho_m} dM  \right]^2 \, \nonumber \\
&~& \times P_0 (k,z) \, ,
\ea
and there are three ingredients which are needed to implement this model: the halo mass function $n$, the linear halo bias $b$, and the halo density 
profile $\rho$ of mean value $\rho_m$. The tree-level matter power spectrum $P_0$ is defined in Eq.~(\ref{eq:p0}); it depends on cosmology through its slope, the growth function, and the transfer function, and it does not include any $\fnl$ correction.
There are many possible ways to parameterise these ingredients and we have 
explored
different alternatives trying to maximise the agreement with the power spectra
measured from the N-body
simulations presented in \citet{Pillepich:2008ka} (hereafter PPH08).
 Concerning the Gaussian part of the halo mass function, we have tried the fitting formula by \citet{Tinker:2008ff}, which is based on dark-matter haloes identified with a SO halo finder in N-body simulations, and the similar function by PPH08, which makes use of a FoF halo finder. For the value of the mean overdensity $\Delta_V$ which is enclosed by the virial radius of haloes, we tried the redshift scaling relation by \citet{Bryan:1997dn} and a constant value of $\Delta_V = 200 $. We also explored the halo exclusion prescriptions \citep{Zheng:2002es,Magliocchetti:2003ee,Tinker:2004gf}. We found that the best agreement with the simulations was recovered using the mass function by PPH08 and $\Delta_V = 200 $, which we will use in the analysis. The agreement is at the $\sim 10 \%$ level as can be seen in Fig.~\ref{fig:test_Pk}.
 For non-Gaussian models, we will then apply the appropriate corrections to the PPH08 halo mass function using the method developed by \citet{LoVerde:2007ri}
and summarised in Eq. (\ref{LVratio}). We will consistently derive the non-Gaussian halo bias using the peak-background split formalism, as reviewed in the Appendix~\ref{sec:appbias}.

We have assumed the analytical description by \citet{Navarro:1995iw,Navarro:1996gj} for the halo density profile of dark matter haloes (NFW). In this approach, the radial density $\rho$ is written in real space as
\be
\rho (r) = \frac {\rho_s} {(r/r_s) \, (1 + r/r_s)^{2}},
\ee
and it is fully specified by the parameters $ r_s, \rho_s$, describing a scaling
radius where the density profile changes its slope and its associated density. 
Alternatively, we can use the concentration $c \equiv R_V / r_s$, where the virial radius $R_V$ is defined as the radius of the sphere whose mean density is $\Delta_V = 200$ times the average density of the Universe.
We can finally write the density profile in Fourier space as \citep{Scoccimarro:2000gm,Rudd:2007zx}
\ba
 \tilde \rho (M, z, k) &=& 4 \pi  \rho_s  r_s^3 \left\{ \sin(k r_s) \left[ \mathrm{Si}(k r_s (1 + c)) -\mathrm{Si}(k r_s) \right] \vphantom{\frac{\sin(c k r_s)}{k r_s (1 + c)}} \right. \\
 &~&- \left. \frac{\sin(c k r_s)}{k r_s (1 + c)} + \cos(k r_s) \left[ \mathrm{Ci}(k r_s (1 + c))- \mathrm{Ci}(k r_s) \right] \right\} \, \nn .
\ea
Here $\mathrm{Si}(x),\mathrm{Ci}(x)$ are the sine and cosine integrals respectively. 
The concentration is known to depend on halo mass and redshift. 
We use the formula by \citet{2011ApJ...740..102K}, obtained from the Bolshoi simulation:
\ba
c (M, z) &=& 9.2 \, \kappa(z) \, D^{1.3}(z) \left( \frac{M}{10^{12} h^{-1} M_{\odot}} \right)^{-0.09} \nonumber \\
&\times& \left[ 1 + 0.013 \left( \frac{M}{10^{12} h^{-1} M_{\odot}} D(z)^{-\frac{1.3}{0.09}}  \right)^{0.25} \right],
\ea
where the function $\kappa(z)$  is introduced to correct for the different definition of the virial radius, and it has values $ \kappa(z=0) = 1.26$, and $ \kappa(z \ge 1) \simeq 0.96 $.  
We have also checked that using the result by \citet{Bullock:1999he} for the concentration does not change significantly the results.
It is worth noticing that the concentration is in principle dependent on the PNG parameter $\fnl$ but, as was shown e.g. by \citet{Smith:2010fh,2011MNRAS.415.1913D}, the effect is very small (at least for local PNG) and
we will ignore it in the rest of this analysis.

\subsection {Power spectrum and comparison with simulations}

With the ingredients described above, we can now calculate the full non-linear matter power spectrum $P_m(k,z)$; 
 to recover the linear regime in the large-scale limit, it is fundamental that 
\be
\int_0^{\infty} n (M, z, \fnl) \, b(M,z,k, \fnl) \, \frac {M} {\rho_m} dM = 1 \:\:\:\:\:\:\:\: (k \to 0) \,.
\ee
We will enforce this constraint in our calculations by adding a constant to the bias in the smallest mass bin, as also described by \citet{Fedeli:2009mt}. This condition should be automatically satisfied, but the need to enforce it explicitly arises due to the binning, the finite integration limits which exist in practice, and other numerical issues.

\begin {figure} 
\begin{center}
\includegraphics[width=0.45\textwidth,angle=0]{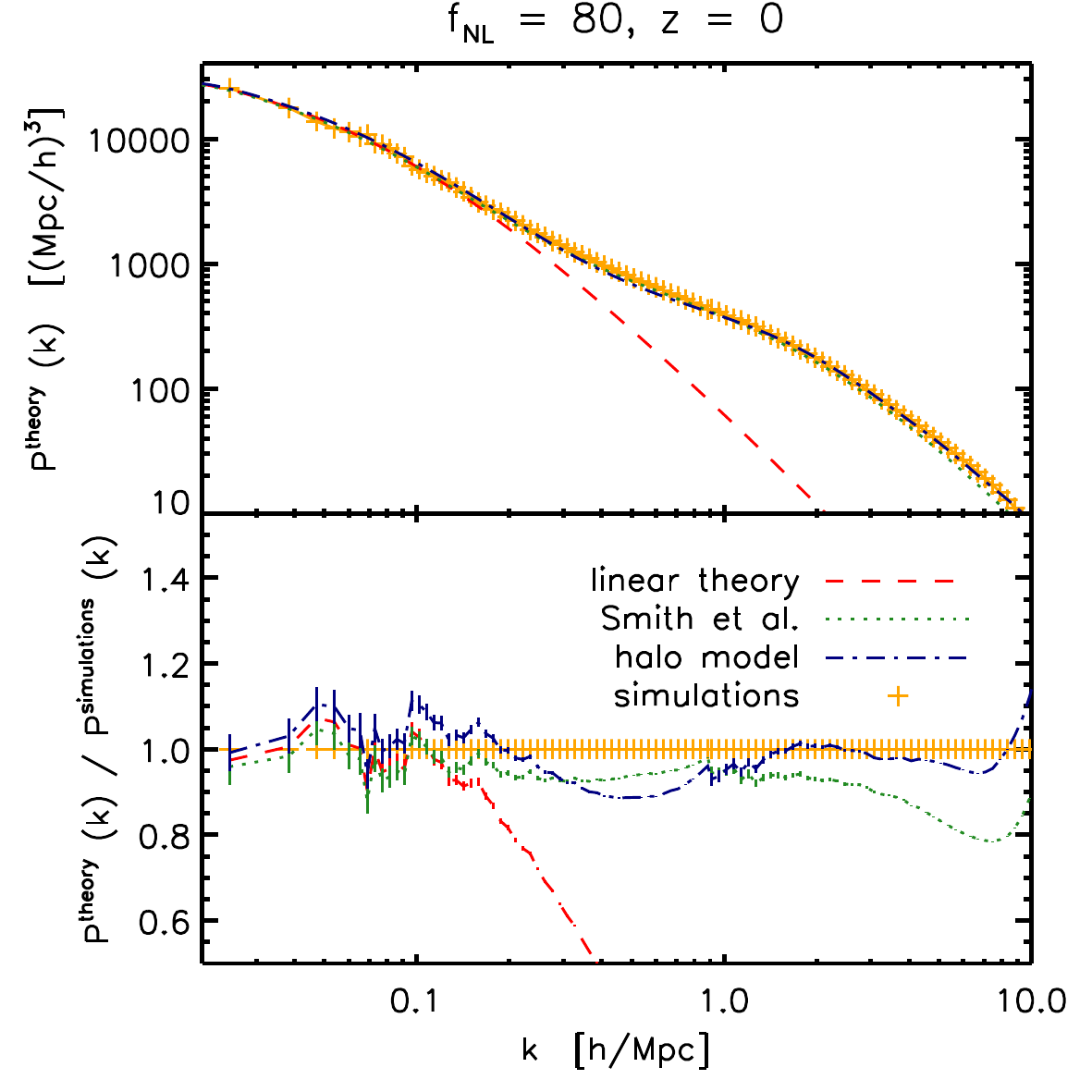}
\end{center}
\caption{Comparison of the theoretical matter power spectra with the measurements from $N-$body simulations by PPH08, with local $\fnl= 80$. The top panel shows the total $P_m(k)$ for the linear (red) and two non-linear models: the halo model (blue) and the model by \citet{Smith:2006ne} (green). Below we plot the ratio $P_m(k)/P_{\mathrm{simulations}}(k)$. All curves are at $z=0$: a similar ($\sim 10 \%$) level of accuracy is obtained at other redshifts.}
\label{fig:test_Pk}
\end{figure}

We then compare the results of the non-linear power spectrum obtained with these recipes with the $N$-body simulations by PPH08. 
Consistently with PPH08, for our calculations we assume the WMAP5 flat \LCDM~ model, with
parameters $h = 0.701$, $\sigma_8 = 0.817$, $n_s = 0.96$, $\Omega_m = 0.279$, $\Omega_b = 0.0462$, $\Omega_{\Lambda} = 0.721$.
 We show the result of the comparisons in Fig.~\ref{fig:test_Pk} at fixed redshift $z=0$, as the comparison results are very similar at different redshifts. In the top panel we can see the full matter power spectra for the choice of local $\fnl = 80$, compared with the simulations. Below we show the residuals $P_m(k)/P_{\mathrm{simulations}}(k)$.
We can see that the agreement is at the $10\%$ level up to scales of $10~h/$Mpc. This is the level of accuracy to be expected from the halo model \citep{Cooray:2002dia}. We can also see a comparison with the method to calculate the non-linear power spectrum using a fit from simulations by \citet{Smith:2006ne}, whose accuracy is of similar order for $k < 2 h/$Mpc, but then degrades further to the $\sim 20 \%$ level.

In Section \ref{sec:haloinacc} we will compare the results of the Fisher matrix analysis obtained using  different models for the non-linear evolution of the matter power spectrum, finding that the level of uncertainty still present in the halo model  does not significantly impact  the results of our forecasts in the galaxy clustering case.

\section {Future galaxy surveys} \label{sec:data}

Wide galaxy surveys stand amongst the most powerful ways of probing cosmology; they can be used to measure several observables, such as tracers of the large-scale structure of the Universe  and indicators of the expansion rate. For example the Sloan Digital Sky Survey (SDSS) \citep{Aihara:2011sj} and its follow-ups have greatly improved our understanding of the standard model, especially thanks to the measurement of galaxy clustering and baryon acoustic oscillations \citep[see e.g.][]{Percival:2007yw}, but also using galaxy clusters \citep{Koester:2007bg} and Type Ia Supernovae \citep{Kessler:2009ys,Lampeitl:2009jq}. Weak lensing surveys \citep[see e.g.][]{Fu:2007qq} are becoming increasingly powerful in constraining cosmology due to the improved control of systematics. The same surveys also provide additional cosmological probes, e.g. their external correlations with the CMB anisotropies \citep{Giannantonio:2008zi}.

All of these observables share part of the information, and will therefore not be fully independent. A considerable challenge for future surveys will be to properly account for their covariance, as was discussed e.g. for lensing and clustering by \citet{Hu:2003pt} and for lensing and galaxy clusters by \citet{Takada:2007fq}.
In this paper we study how the constraining power on cosmology will improve in the near future using two examples of upcoming surveys, focussing on the Dark Energy Survey (DES) \citep{Abbott:2005bi} for the DETF stage III and an Euclid-like survey \citep{2009arXiv0912.0914L} for stage IV. It is reasonable to expect that other surveys in the same class will have comparable performances, such as in particular the American proposed WFIRST class IV mission.

\subsection {Stage III: the Dark Energy Survey}

The Dark Energy Survey (DES)\footnote{\texttt{http://www.darkenergysurvey.org/}} \citep{Abbott:2005bi} is an optical and near-infrared survey which is currently being deployed at the Cerro Tololo observatory in the Chilean Andes, and  has recently come to light in late 2011. The wide-field survey will cover 5,000 sq. deg. in the Southern sky, reaching $\sim 24$ magnitude in the SDSS bands $g, r, i, z$ and $Y$. Additionally, the $J,H,K$ bands for the same fields are expected to be added from the ESO Vista survey. About 300 million galaxies are expected to be observed with shapes, photo-z's and positions, as well as 100,000 galaxy clusters and 1,000 Type Ia SNe using a smaller repeated imaging survey.

In this paper we will study the Fisher-matrix forecasted errors on dark energy and PNG using the weak lensing, galaxy clustering, and the combined data from this survey. For this purpose, we will assume the specifications summarised in Table \ref{tab:techspec} and
that the redshift distribution of the sources can be well approximated by the law  \citep{1994MNRAS.270..245S} 
\be
\frac{dN}{dz}(z) = \frac{1}{\Gamma \left(\frac {\alpha + 1} {\beta} \right)} \, \beta \, \frac {z^{\alpha}}{z_0^{\alpha+1}} \, \exp \left[ - \left( \frac{z}{z_0} \right)^{\beta} \right],
\ee
where the parameters are set to $\alpha = 2$, $\beta = 1.5$,
in which case $z_0$ is related to the median redshift by $z_0 \simeq \bar z / 1.41$. We will finally split this distribution in eight equally populated redshift bins. Due to the uncertainty in the photo-z determination, we will write the theoretical redshift distribution of the sources as the convolution of these redshift bins with a Gaussian characterised by a dispersion matching the photometric redshift uncertainty $\sigma_z(z)$.
This approach assumes that the photometric redshift estimation is perfectly calibrated to a
spectroscopic sample; the distribution of the redshift errors is Gaussian and the r.m.s. error
$\sigma_z (z)$ is known; catastrophic errors can be identified and downweighted in the analysis.
See \citet{2008MNRAS.389..173K,2010MNRAS.401.1399B} for more detailed approaches which take into account more of the possible systematics arising in this case.

Finally, in the absence of a detailed model we will assume the fiducial value of the galaxy bias to follow the law $b(z) = \sqrt{1+z}$ following \citet{Rassat:2008ja}, but we will study the effect of modifying the fiducial bias in Section~\ref{sec:bchange}.

\begin {table*}
\begin {center}
\begin{tabular}{c c c c c}
\hline
Parameter     & description             &  Euclid  photometric   &  Euclid spectroscopic &  DES \\
\hline
$\sigma_z (z)/(1+z)$  & redshift uncertainty    & 0.05            & 0.001  & 0.1 \\
$\bar z$      &median redshift          &  1.0                &  1.1  &  0.8 \\
$n$        &   galaxy density           &  40    arcmin$^{-2}$            & 0.7775   arcmin$^{-2}$    &  12   arcmin$^{-2}$    \\
$A$       & surveyed area             & 20 000  sq deg            & 20 000  sq deg  & 5 000 sq deg  \\
$\gamma$ & intrinsic ellipticity noise    &  0.247           &  ---  & 0.16 \\
$dN/dz (z)$ & galaxy distribution            & Smail et al.     & Geach et al.  & Smail et al. \\
 & parameters of gal. dist.       & $\alpha = 2 $           &  flux cut = $4 \cdot 10^{-16}$ erg/s/cm$^2$   &$\alpha = 2 $   \\
&                                             & $\beta = 1.5 $         & efficiency = 35\%  & $\beta = 1.5 $       \\
$ M$ & number of redshift bins          & 12               & 12  &  8 \\
\hline
\end{tabular}
\caption{Technical specifications of the surveys used for forecasts.}
\label{tab:techspec}
\end{center}
\end {table*}

\subsection {Stage IV: Euclid}

Euclid\footnote{\texttt{http://sci.esa.int/euclid}} is a future space mission of the European Space Agency (ESA), which is expected to survey the whole extragalactic sky (up to 20,000 sq. deg.) from the L2 point in space  \citep{2009arXiv0912.0914L}. Launch is currently scheduled in 2018.  In our analysis we approximate this planned survey with the settings and specifications described in the Euclid Assessment Study Report \citep{2009arXiv0912.0914L}.
Although the specifications of the survey have since evolved  \citep{RedBook}, this does not affect the current study because our intention is to draw broad conclusions that will be relevant to the whole class of stage IV surveys including WFIRST, LSST, and others. See in any case Section~\ref{sec:redbook} for a discussion of how the results change when using the latest Red Book specifications.
The Euclid mission is expected to perform two main surveys, photometric and spectroscopic.

\begin{figure} 
\begin{center}
\includegraphics[width=0.45\textwidth,angle=0]{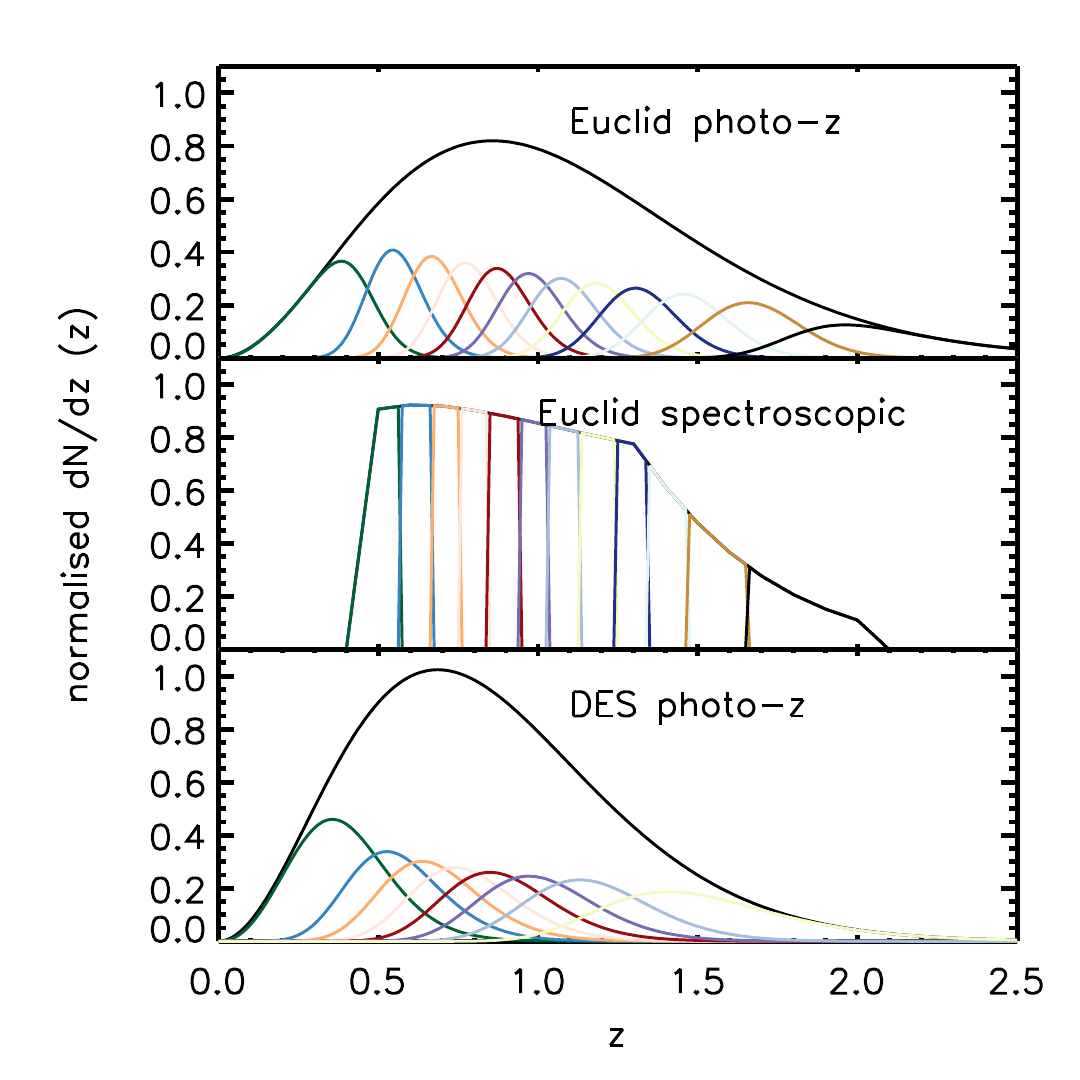}
\end{center}
\caption{Redshift distributions used for the forecasts of both photometric and spectroscopic data sets for the Euclid satellite and the DES. The photo-z distributions are given by an analytic function \citep{1994MNRAS.270..245S}, while the spectroscopic part is numerically estimated by \citet{2010MNRAS.402.1330G}, where the Euclid specified flux cut is used, $4 \cdot 10^{-16} $ erg s$^{-1}$ cm$^{-2}$. 
The distributions have been already convolved with the probability density
function of redshift measurement errors.
}
\label{fig:dNdz_Pk}
\end{figure}

 The photometric part should measure photo-z and ellipticities in the optical and near-infrared bands (one broad visual band $R$+$I$+$Z$ and $Y, J, H$ IR bands), up to mag $\sim 24.5$ in the visual and 24 in the IR. The requirement specifications are described  in Table \ref{tab:techspec}.
 The expected number of observed galaxies is of the order of a billion. For this survey we will also use the approximated redshift distribution by \citet{1994MNRAS.270..245S}, dividing the sample in 12 redshift bins, whose distribution is again convolved with the expected photometric errors, as shown in Fig.~\ref{fig:dNdz_Pk}. The photometric galaxies are distributed in the redshift range $0<z<2.5$.

The spectroscopic survey is expected to use a slitless spectrometer which will mainly detect the H$\alpha$ emission line of galaxies. The spectrometer will have a resolution $\lambda/\Delta \lambda = 500$, giving a redshift uncertainty of $\sigma_z(z) = 0.001 (1+z)$. The wavelength range of this instrument will be limited to  1000 nm $< \lambda <$ 2000 nm, meaning that only galaxies at $0.5 < z < 2$ will have measurable H$\alpha$ lines and thus redshifts. The limiting flux is placed at $4 \cdot 10^{-16} $~erg~s$^{-1}$~cm$^{-2}$, which combined with the expected success rate of the spectrometer $e = 35\%$ yields $\sim 60$ million galaxies, using the predicted tabulated calculations by \citet{2010MNRAS.402.1330G}, which was based on empirical
 data of the luminosity function of H$\alpha$ emitter galaxies out to $z=2$. We will use this tabulated prediction as our fiducial redshift distribution of the sources, consistently with the Euclid Study Report specifications \citep{2009arXiv0912.0914L}. The remaining specifications are again to be found in Table~\ref{tab:techspec}.
 We will finally split the distribution into 12 equally populated redshift bins, as we can see in Fig.~\ref{fig:dNdz_Pk}.
 Again, we will take the fiducial value of the galactic bias to follow the law $b(z) = \sqrt{1+z}$ following \citet{Rassat:2008ja}, which is a good approximation to recent studies from semianalytic models of galaxy formation such as e.g. by \citet{2010MNRAS.405.1006O}, but we will study the effect of changing this choice in Section~\ref{sec:bchange}.

\section {Fisher matrix forecasts}  \label {sec:forec}

\subsection {Formalism}
Let us assume, following e.g. \citet{Tegmark:1996bz,Tegmark:1997rp}, that we have a data vector $\mathbf x$, whose elements are random variables which depend
on the cosmological parameters $\boldsymbol {\vartheta}$, with a likelihood function $L(\mathbf x, \boldsymbol \vartheta)$. Then, if we fix $\mathcal L \equiv - \ln L$, the Fisher matrix can be defined as the ensemble average
\be
\mathbf F_{ij} \equiv  \left\langle \frac  {\partial^2 \mathcal L}{\partial \vartheta_i \, \partial \vartheta_j}  \right\rangle \, ,
\ee
and the Cram\'er-Rao inequality proves that its inverse represents the best possible covariance matrix for the measurement of the parameters $\boldsymbol \vartheta$ with the help of unbiased estimators. In the limit of large data sets, if the distribution becomes Gaussian, this inequality becomes an equality, and the standard deviation on a given parameter $\vartheta_i$ is given by $\sigma (\vartheta_i) = \sqrt {(\mathbf F^{-1})_{ii}}$.
Under the assumption of a Gaussian likelihood function $L$, for data of mean $\boldsymbol \mu \equiv \langle \mathbf x \rangle$ and covariance matrix $\boldsymbol {\Sigma} \equiv \langle \mathbf x \mathbf x^T \rangle - \boldsymbol \mu \boldsymbol \mu^T$, the Fisher matrix can be written as the sum of two pieces:
\be \label{eq:fisherdef}
\mathbf F_{ij} = \frac 1 2 \mathrm {Tr} \left[\boldsymbol {\Sigma}^{-1} \, \boldsymbol {\Sigma}_{,i} \, \boldsymbol {\Sigma}^{-1} \, \boldsymbol {\Sigma}_{,j}   \right] + \boldsymbol {\mu}_{,i}^T \, \boldsymbol {\Sigma}^{-1} \, \boldsymbol {\mu}_{,j},
\ee
where the commas denote derivatives with respect to the parameters.

 To calculate the Fisher matrix, we need first to define our parameter set and its fiducial model, which we take in our base analysis to be for the galaxy clustering case: $\boldsymbol{\vartheta_0} = \{ \Omega_{\Lambda} = 0.721, \, \Omega_b = 0.0462,  \, \Omega_m = 0.279,  \, h = 0.701,  \, n_s = 0.96,  \, \sigma_8 = 0.817,  \, w_0 = -1.0,  \, w_a = 0,  \, \fnl = 0,  \, \mathbf b = \mathbf b_{\mathrm{fid}}  \}$, where the usual set of cosmological parameters is chosen to match the WMAP5 set used in \citet{Pillepich:2008ka}, the dark energy equation of state is parameterised as $w(z) = w_0 + w_a \, z / (1 + z) $ \citep{Chevallier:2000qy}, and the array $\mathbf b $ of fiducial value $ \mathbf b_{\mathrm{fid}}$ is a set of nuisance parameters to describe the total scale-independent part of the galaxy bias in each redshift bin (see discussion in Section \ref{sec:lssandprim}). Notice that the degeneracy between $\sigma_8$ and $b$ is broken by the non-linear corrections to the power spectrum at $k>0.1 h/$Mpc. For the lensing case, as there is no biasing, we do not include bias nuisance parameters.
We will also consider different parameter arrays and extra nuisance parameters in some cases, as described in the following sections.

\subsection{Observables} \label{sec:observables}

Starting from a 3D random field $ \varphi (\mathbf{x})$, its 3D power spectrum $P (k)$ can be defined as in Eq.~(\ref{eq:powerspecPhi}).
Similarly, if we consider a 2D random field on a sphere $\xi(\hat \mathbf n)$, the angular power spectrum $C_l$ is given by $\langle a_{lm} \, a_{l'm'}^*   \rangle  =  \delta_{ll'} \delta_{mm'} \, C_l $,
where $a_{lm}$ are the spherical harmonic coefficients of the field.
We will use in our analysis the information contained in all the relevant two-point correlation functions (or power spectra) which are measurable by the the DES and Euclid missions: we shall consider the 2D power spectra of weak lensing (using the photometric redshift catalogue) and galaxy clustering (using both the photometric and spectroscopic catalogues), and the 3D power spectrum for clustering (spectroscopic), as described below. We will first present the forecasts for each single probe, and finally combine lensing and clustering, including all the relevant two-point correlations and covariances.
For all numerical calculations we use the iCosmo package by \citet{Refregier:2008fn}.

An important point is whether this basic Fisher setting, which assumes a Gaussian distribution of the covariance, is accurate enough. More extended calculations have shown that extra terms arise, due to the connected matter trispectrum and also to the coupling of large-scale modes which are outside the volume of the survey \citep{Sato:2009ct,Takada:2008fn}. The effect of these terms is to reduce the signal-to-noise ratio obtained from the small scales, thus mostly affecting the cosmic shear results. The worsening of the lensing marginalised errors is expected in most cases not to exceed the $\sim 10 \%$ level \citep{Takada:2008fn} and thus, while it will be important for data analysis,  we will ignore it at the forecasting stage.

\subsection {Cosmic shear}

We can describe a tomographic galaxy survey of $N$ galaxies in $ M$ redshift bins with a set of distribution functions for the galaxy density fluctuations in each $i$-th bin $dN_i/dz (z)$, so that the total number of galaxies in the bin is
\be
N_i = \int_0^{\infty} dz' \, \frac {dN_i}{dz'} (z') \:, \hspace{2cm} N = \sum_{i=1}^M N_i \,.
\ee
In a Friedmann-Robertson-Walker universe of total energy density $\Omega_0$ we can write the curvature of the hypersurfaces at constant cosmic time as
\be
K = \frac{H_0^2}{c^2} (\Omega_0 - 1)  \, ,
\ee
and the comoving angular diameter distance is
\be
r_K (r) = \left\{ \begin{array}{rl}
 K^{-1/2} \sin (K^{1/2} r) &\mbox{ if $K>0$} \\
 r &\mbox{ if $K=0$} \\
 (-K)^{-1/2} \sinh (-K^{1/2} r) &\mbox{ if $K<0$} \\
      \end{array} \right. \, ,
\ee
which coincides with the comoving distance $r$ in the flat case.
 The spatial part of scalar metric perturbations can be decomposed in terms of the solutions of the Helmholtz differential equation \citep{Abbott:1986ct} 
\be
\left( D^2 + k^2 \right) Q (\mathbf{r})  = 0  \, ,
\ee
where $D^2$ is the covariant Laplacian in curved space. The eigenfunctions $Q(r)$ generalise the concept of plane waves to curved spacetime. Introducing spherical harmonics leads to variable separation and the radial part of the eigenfunctions
can be written as $R_{k,{l}} (r) = \Phi_{\beta}^{l} (r)$ \citep{Abbott:1986ct},
$\beta = \sqrt {k^2 + K} $ , and $\Phi_{{l}}^{\beta}(r)$ are the 
ultra-spherical Bessel functions 
(note that \cite{Abbott:1986ct} write their explicit form in terms
of a different radial coordinate).

We can use the weighting kernels for cosmic shear \citep{Bartelmann:1999yn}
\be
W^{\epsilon_i}(z) = \frac{1}{N_i} \,  \frac {3 H_0^2 \Omega_m}{2 \, c^2 \, a (z)}  \,\int_z^{\infty} dz' \, \frac {dN_i}{dz'} (z') \,  \frac  {r_K [r (z') - r(z) ]  }{ r_K [ r(z') ]  }  \, 
\ee
to calculate the projected 2D cosmic shear power spectrum between the bins $i,j$ as 
\ba
 C^{\epsilon_i \epsilon_j}_{{l}} &=& \frac{2}{\pi} \int_0^{\infty}   d \beta \, \beta^2 \, \int_0^{\infty}  dr_1 \,W^{\epsilon_i} (r_1) \, \Phi_{{l}}^{\beta}(r_1)  \nonumber \\
&~&  \cdot \, \int_0^{\infty} 
dr_2 \,
W^{\epsilon_j}(r_2) \, \Phi_{{l}}^{\beta}(r_2) \, \cdot \,  P_{m} (\beta, r_1, r_2) \, ,
\ea
where $P_m$ here denotes the cross-spectrum between the matter distribution at two different cosmic epochs corresponding,  on our past light cone, to distances $r_1$ and $r_2$. \footnote{For a universe with positive curvature the integral
over the corrected wavenumber $\beta$ should be replaced by a discrete
sum such that $\beta \geq 3$ and $\beta>l$.}
This can then be written in the Limber approximation as
\be \label{eq:Limber}
 C^{\epsilon_i \epsilon_j}_{{l}}  \simeq \int_0^{\infty} \frac {d r}{r_K^2(r)} \, W^{\epsilon_i}[z(r)] \, W^{\epsilon_j}[z(r)] \, P_{m} \left[ k = \frac {{l} + \frac 1 2} {r_K(r)}; z(r) \right] \, ,
\ee
and the full 3D matter power spectrum can be calculated e.g. using the halo model formalism of Section~\ref{sec:nonlin}.

The observed lensing power spectra $\tilde C^{\epsilon_i \epsilon_j}_{{l}}$ may be modelled as a sum of the theoretical spectra and the noise. For each pair of redshift bins $i,j$, we have
\be
  \tilde C^{\epsilon_i \epsilon_j}_{{l}}  = C^{\epsilon_i \epsilon_j}_{{l}} +  N^{\epsilon_i \epsilon_j}_{{l}} \, ,
\ee
where the noise term $N^{\epsilon_i \epsilon_j}_l$ is due to the shape noise
\ba
N_{{l}}^{\epsilon_i \epsilon_j} &=& \delta_{ij} \frac {\gamma^2} {\bar n_{i}} \, . 
\ea
Here $\bar n_{i}$ is the angular number density in the $i-$th bin, and $\gamma$ is the r.m.s. shear arising from the intrinsic ellipticities of the galaxies, which is a specification of the survey (see Table~\ref{tab:techspec}).

Another important source or errors in cosmic shear are intrinsic alignments: we do expect a fraction of neighbouring galaxies will have correlated ellipticities due to tidal fields. This effect introduces a systematic into the cosmic shear analysis which has to be taken into account when constraining cosmology from real data \citep{Joachimi:2009ez, Kirk:2010zk}, while we will neglect it for our forecast as the biasing introduced on the parameter estimation typically does not exceed the $\sim 10 \% $ level.

We can now calculate the Fisher matrix element from Eq.~(\ref{eq:fisherdef}) for the case of weak lensing. A common assumption in this case is to identify the observables $\mathbf x$ with the $a_{{l} m}^{\epsilon_i \epsilon_j}$  instead of the $C_l^{\epsilon_i \epsilon_j}$, because the $a_{{l} m}$ are at least in the linear regime Gaussian. This brings in the advantage of cancelling the second term of Eq.~(\ref{eq:fisherdef}), since $ \boldsymbol {\mu_i} = \langle a^i_{{l} m} \rangle = 0 $, $\, \forall \, \{i,{l},m\}$. By definition, the variance of each of the $(2{l} + 1)$ $a_{{l} m}^{\epsilon_i \epsilon_j}$ is given by  $\boldsymbol{\Sigma}_{ab} = C_a^{\epsilon_i \epsilon_j} \delta_{ab}$, and thus Eq.~(\ref{eq:fisherdef}) yields \citep{Hu:2003pt, Amara:2006kp}
\be \label {eq:fish2d}
F^{\epsilon}_{\alpha \beta} = f_{\mathrm{sky}} \sum_{{l}={l}_{\min}}^{{l}_{\max}} \frac{(2 {l} + 1) }{2} \, \mathrm{Tr} \left[ \mathbf D^{\epsilon}_{{l} \alpha} \,  \left( \tilde \mathbf C^{\epsilon}_{{l}} \right)^{-1} \, \mathbf D^{\epsilon}_{{l} \beta} \,  \left( \tilde \mathbf C^{\epsilon}_{{l}} \right)^{-1} \right],
\ee
where $f_{\mathrm{sky}}$ is the sky coverage of the survey. Here $\tilde \mathbf C^{\epsilon}_{{l}}$ is a matrix of dimensions $ M \times  M $, whose elements are the lensing power spectra at a fixed ${l}$ between each pair of redshift bins $\tilde C_{{l}}^{\epsilon_i \epsilon_j}$.  Finally, $ \mathbf D^{\epsilon}_{{l} \alpha}$ is a matrix containing the derivatives of the spectra with respect to each cosmological parameter $\vartheta_{\alpha}$, whose elements are defined as
\be
D^{\epsilon_i \epsilon_j}_{{l} \alpha} = \frac{\partial C^{\epsilon_i \epsilon_j}_{{l}} }{\partial \vartheta_{\alpha}} \, .
\ee
We shall calculate the forecasted weak lensing power spectra accordingly with the proposed specifications of the DES and Euclid missions, as described above.

The range of multipoles used in the sum of Eq.~(\ref{eq:fish2d}) $[{l}_{\min}, {l}_{\max}]$ is a sensitive issue, since we would like to set it as broad as possible to use all available information.
The value of ${l}_{\max}$ is related to how far we can trust our modelling of the non-linear regime.  In the case of weak lensing, as we are free from the problems of non-linear biasing, we take ${l}_{\max} = 20,000$, which gives $k^{\epsilon}_{\max} \simeq 8.5 h/$Mpc at $z=1$ assuming the best-fit WMAP cosmology. Again this assumes that the effect of baryons to the matter power spectrum can be either ignored or accurately modelled.
Another issue which may affect the analysis at these small scales is the non-Gaussian contribution to the covariance; we do not account for these corrections whose effects are expected to be small, as discussed at the end of Section \ref{sec:observables}.
 Further, the theory calculation in this regime is strongly dependent on the parameters of the halo model, such as the concentration, the subhalo distribution etc.
We will discuss in Section~\ref{sec:smallscales} the effects of changing this limit, where we will find that  with  more conservative choices  the differences are still small (see Fig.~\ref{fig:lmax}): for example, the marginalised error on $\fnl$ including Planck priors degrades in this case by only $15\%$ between ${l}_{\max} = 20,000$ and ${l}_{\max} = 5,000$.

\begin {figure} 
\begin{center}
\includegraphics[width=0.45\textwidth,angle=0]{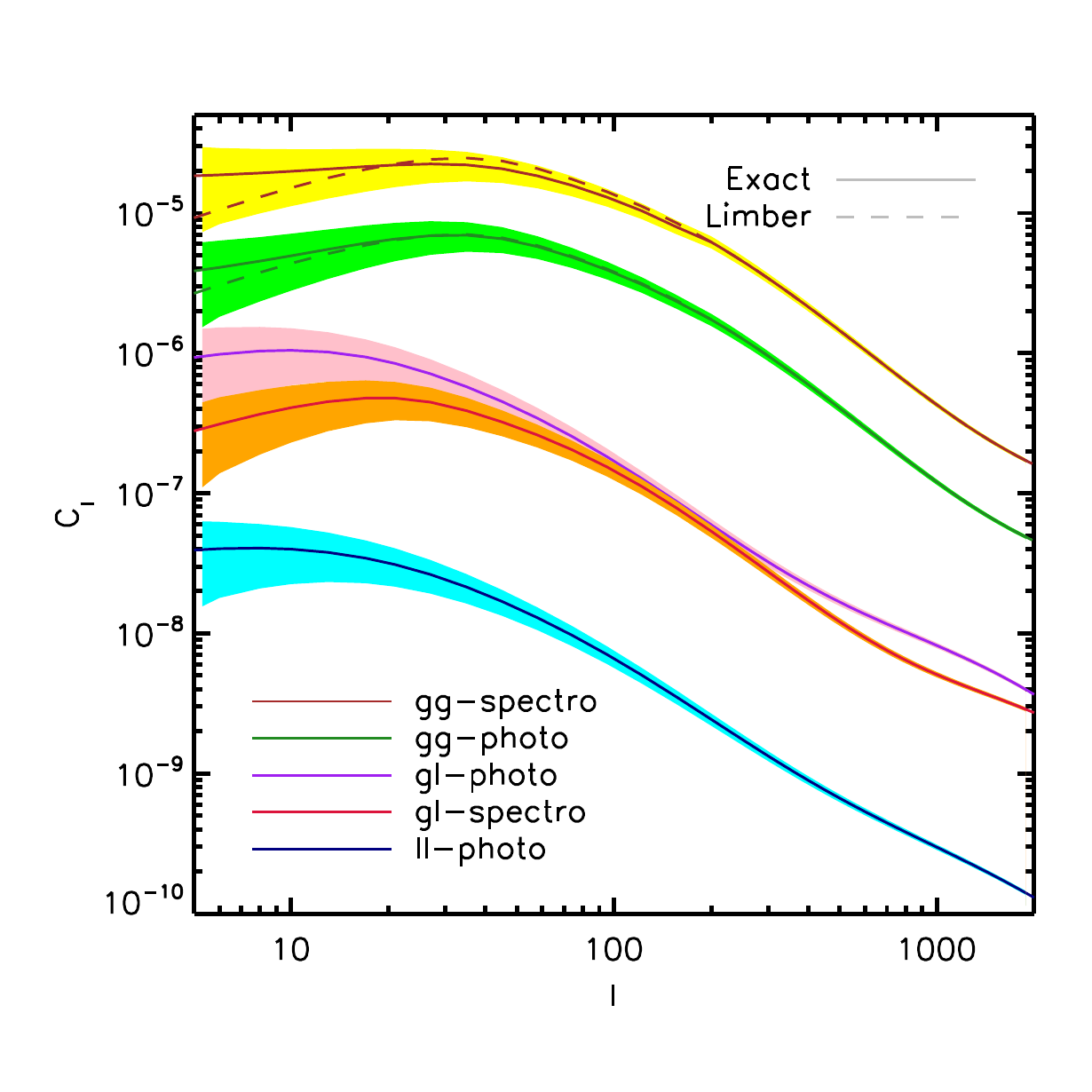}
\includegraphics[width=0.45\textwidth,angle=0]{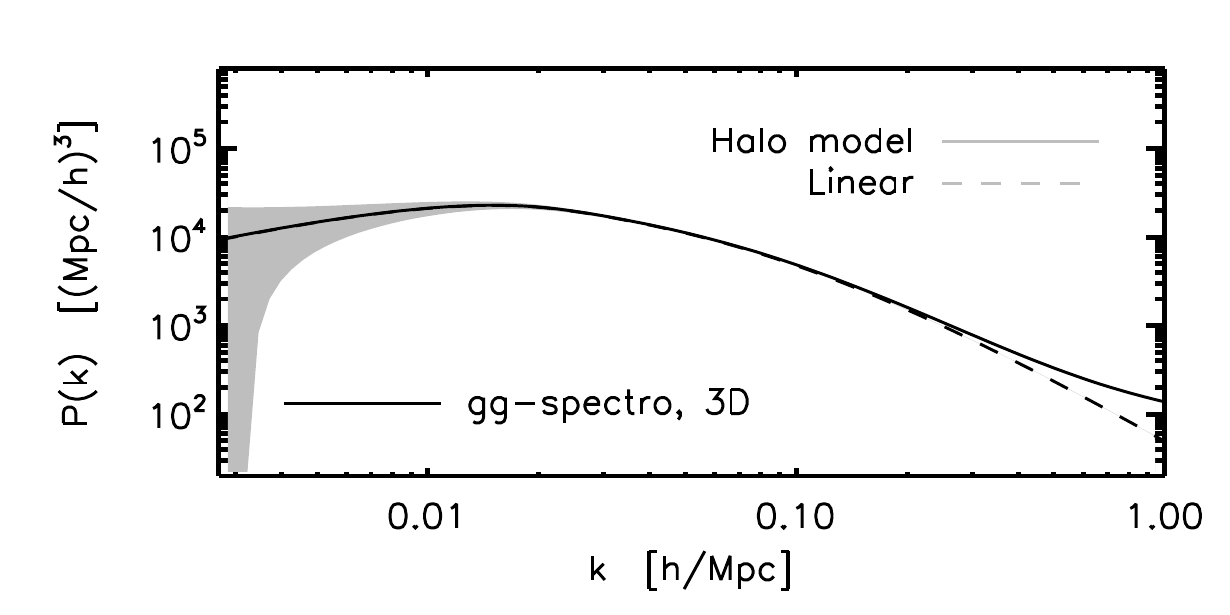}
\end{center}
\caption{Summary of all observables used in this work for the fiducial model described in Section \ref{sec:forec}. In the upper panel, we show the projected 2D spectra for lensing, galaxy clustering, and their cross-spectra for photometric and spectroscopic surveys. In the bottom panel we show the 3D galaxy power spectrum. Note that the Limber approximation is inaccurate on large scales for the galaxy-galaxy spectra, as well as the linear-theory power spectrum is on the small scales (dashed lines). For the lensing and galaxy-lensing cases, the Limber approximation works well due to the wider distribution of the sources. The shaded areas represent cosmic-variance errors for a half-sky survey. A redshift bin centred around $z = 1$ is used in all cases.}
\label{fig:Cls}
\end{figure}

The minimum multipole used is also important. It is expected that the accuracy of the Limber approximation will deteriorate for $l \to 0$.
Here, to be complete and conservative, we will only use this approximation at small scales, for ${l}  \ge 200$, and will use the complete exact formula at larger scales (using the linear approximation for the cross spectrum at two times). In this way, we can extend the calculation all the way to the largest scales, and will take ${l}_{\min} = 5$.
However, as shown in Fig.~\ref{fig:Cls}, the exact calculation is important only for the case of galaxy clustering, where it can not be neglected as in the presence of non-Gaussianity, the large scales (small multipoles) are the most affected by the scale-dependent bias.
Note that the Limber approximation overestimates the power on the largest scales
for positive $\fnl$ while it underestimates it for a Gaussian model. Since the Fisher matrix is computed taking the derivatives of the signal with respect to 
model parameters, the Limber approximation is rather inaccurate for a fiducial
with $\fnl=0$.

We can see the resulting 2D marginalised forecasts for a Euclid-like survey for local PNG in Fig. \ref{fig:fish_single}, and all the marginalised 1D error bars in Fig.~\ref{fig:all1D}. The results are also summarised in detail in Tables~\ref{tab:bigtable}, \ref{tab:bigtable_ort}, \ref{tab:bigtable_equi} for the three PNG configurations. We present the results for DES in Table~\ref{tab:bigtableDES} for the local case.

\subsection {Galaxy clustering, 3D}

The second observable we consider is the clustering of galaxies, which is typically measured via the 3D power spectrum using the spectroscopic part of the survey.
In this case we then consider measurements of the galaxy power spectrum for each $i$-th redshift bin $P_{g_i}(k) = b_{g_i}^2 (k) \, P_{m}(k)$, where $b_{g_i} (k) = \left[ b_{g_i} + \Delta b_i(k) \right] $ is the galaxy bias of the $i$-th bin.
The scale-dependent part of the bias is computed as explained in Section 
\ref{sec:NG} while we consider the scale-independent part as 
a nuisance parameter to be marginalised over.
Due to the high accuracy of the spectroscopic redshifts, we can ignore any cross-spectra between different redshift bins, since the redshift distribution functions will have negligible overlap, and we will also neglect the effect of covariances due to the longitudinal modes. We also include corrections due to two effects, as described e.g. by \citet{Seo:2003pu, Song:2008qt}: the redshift-space distortions and the Alcock-Paczynski effect.

Redshift-space distortions are caused by the peculiar motion of galaxies \citep{Kaiser:1987qv}, and produce an enhancement of the overdensity field $\delta$ proportional to $(1+f\mu^2)$, where $f$ is the logarithmic derivative of the growth function with respect to the expansion factor $a$, and $\mu$ is the cosine of the angle to the line of sight. If we can exclude velocity bias, the correction of the power spectrum becomes (see e.g. \citet{Peacock:1993xg})
\be \label{eq:FoG}
P^s_{g_i}(\mathbf{k}) = P_{g_i}(\mathbf{k}) \left[ 1 + \mu^2
  \beta_{g_i} (k) \right]^2 \, F\left( \frac {k^2 \mu^2 \sigma_{v, i}^2} {H^2(z)} \right) \,
\ee
where $\beta_{g_i}(k) = f / b_{g_i}(k)$, $\sigma_{v, i}$ is the 1D pairwise velocity dispersion of the galaxies, and $H(z)$ is the Hubble function.
We will use for the function $F $ the form $F(x) = e^{-x}$ \citep{Percival:2007yw}. This gives a small correction whose cosmological content
can be ignored and treated as a nuisance parameter to be marginalised
over \citep{Song:2008qt}. In more detail, we can decompose the velocity
dispersion in each redshift bin $i$ as the sum in quadrature of an
intrinsic term coming from the finger-of-god (FoG) effect,
$\sigma_{\mathrm{FoG}, i}$, and a term due to the redshift uncertainty of the survey 
$\sigma_z$ given in Table~\ref{tab:techspec}, as
\be
\sigma_{v, i}^2 = (1+z)^2 \left[ \frac{\sigma_{\mathrm{FoG}, i}^2}{2} +  c^2 \sigma_z^2 \right] \, .
\ee
Since the redshift evolution of this intrinsic term is uncertain, it would be desirable to introduce independent parameters $\sigma_{\mathrm{FoG}, i}$ for each redshift bin, and then marginalise over them. However, we found that, given that this factor enters the calculation together with the function $1/H(z)$ in Eq.~(\ref{eq:FoG}), leaving these parameters free to change brings in severe degeneracies with any other cosmological parameter which is able to alter $H(z)$. For this reason, we have been forced to adopt a less conservative approach, where we choose a functional form
\be
\sigma_{\mathrm{FoG}} (z) = \sigma_{\mathrm{FoG}, 0} \, \sqrt{1+z} \,
\ee
and only marginalise on one parameter $\sigma_{\mathrm{FoG}, 0}$. This scaling
with redshift assumes that the observed galaxies reside in halos with nearly
constant mass.
It is also possible to make this approach more conservative, e.g. by considering more complicated parameterisations for this function, but again the constraints on all parameters which enter the Hubble expansion $H(z)$ will worsen dramatically.
As a fiducial value, we take $\sigma_{\mathrm{FoG}, 0} = 250 $ km/s, as the Euclid
spectroscopic sample is made of star-forming galaxies, which are not generally located in massive haloes, and are thus expected to have a low velocity dispersion.

On the other hand, the Alcock-Paczynski effect occurs since  to infer galaxy distances from the observed redshifts and positions we have to use a reference cosmology, which is different from the `true' one. This can be corrected, and the true values are \citep{Seo:2003pu}
\ba
k_{\mathrm{true}} &=& \left[ k_{\mathrm{ref}}^2 \, (1 - \mu_{\mathrm{ref}}^2)    \frac {D^2_{\mathrm{ref}}(z)} { D^2_{\mathrm{true}}(z)}   + (k_{\mathrm{ref}} \, \mu_{\mathrm{ref}})^2   \frac{H^2_{\mathrm{true}}(z)} {H^2_{\mathrm{ref}}(z)}   \right]^{1/2} \nonumber \\              
\mu_{\mathrm{true}} &=& k_{\mathrm{ref}} \, \mu_{\mathrm{ref}} \,\frac{H^2_{\mathrm{true}}(z)} {H^2_{\mathrm{ref}}(z)} \, \frac{1}{k_{\mathrm{true}}} \, .
\ea
Here we will identify the reference cosmology with our fiducial model for simplicity.
Finally, the observable corrected power spectrum, which we will use to calculate the Fisher matrix, is given by
\ba
\tilde P_{g_i} (k_{\mathrm{true}}, \mu_{\mathrm{true}}) &=& \frac {D^2_{\mathrm{true}} (z) \, H_{\mathrm{ref}}(z) } {D^2_{\mathrm{ref}} (z) \, H_{\mathrm{true}}(z)} \, b_{g_i}^2 (k_{\mathrm{true}}) \left[ 1 + \beta_{g_i} (k_{\mathrm{true}}) \, \mu^2_{\mathrm{true}}  \right]^2 \nonumber \\
&~&\times \,  P_m (k_{\mathrm{true}}) \, F\left( \frac {k_{\mathrm{true}}^2 \mu_{\mathrm{true}}^2 \sigma_{v, i}^2} {H^2(z)} \right)   + P_{\mathrm{shot}, g_i}.
\ea
The last term on the r.h.s. is due to shot noise. If this is Poissonian, its
fiducial value is $P_{\mathrm{shot}, g_i} = 1 / \bar n_i $. However, it is possible to be more conservative and assume that the mean galaxy density is not perfectly known, or that there are other sources of shot noise.
We will include these unknown contributions by introducing one additional nuisance parameter for it in each bin, and marginalising over them.

The Fisher matrix calculation is here based on the works by \citet{Feldman:1993ky,Tegmark:1997rp} amongst others. In the 3D case, it is customary to identify each of the observables $\mathbf x_i$ with the average power in a thin shell of radius $k_i$ in Fourier space, of width $dk_i$ and volume $V_i = 4\pi k_i^2 dk_i / (2 \pi)^3 $. In this case, for each redshift bin $a$ and angle $\mu_b$ we have non-zero means $\boldsymbol \mu_i \simeq \tilde P_{g_a}(k_i,\mu_b)$, and covariances
\be
\boldsymbol{\Sigma}_{ij} (\mu_b) \simeq 2 \frac {\tilde P_{g_a}(k_i, \mu_b) \tilde P_{g_a}(k_j, \mu_b)}{V_i V_{\mathrm{eff}} (k_i, \mu_b)} \delta_{ij} \, ,
\ee
where the effective volume is
\be
V_{\mathrm{eff}} (k, \mu) = \int \left[ \frac{\bar n (\mathbf{r}) \tilde P_{g_a}(k, \mu)} {1 + \bar n(\mathbf{r}) \tilde P_{g_a}(k, \mu)}   \right]^2 d^3 r \, ,
\ee
and $n(\mathbf{r})$ is the selection function of the survey.
The dominant term of the Fisher matrix is now the second one in Eq.~(\ref{eq:fisherdef}) which, adding the effects of redshift-space distortions, can be finally written as \citep{Seo:2003pu}
\be
F^{\mathrm{3D}}_{\alpha \beta} =  \int_{-1}^1  \int_{k_{\min}}^{k_{\max}} \frac {\partial \ln \tilde P (k, \mu)} {\partial \vartheta_{\alpha}} \frac {\partial \ln \tilde P(k, \mu)} {\partial \vartheta_{\beta}} \, V_{\mathrm{eff}} (k, \mu) \, \frac{\pi k^2}{(2\pi)^3} \, d k \, d\mu \, .
\ee
Since here we are not accounting for bias non-linearities nor scale-dependences at small scales, our calculated galaxy power spectrum is exact only in the linear regime; therefore 
to limit the effects of non-linearities, in particular for what concerns galaxy biasing (see \citet{2011MNRAS.415..829R}), we cut our calculation at a scale $k_{\max} = 0.15 \, h/$Mpc at $z=0$.
We can then make this limit more meaningful by evolving it as a function of $z$, by imposing the condition
\be \label{eq:var}
\sigma^2(z) = \int_{k_{\min}}^{k_{\max} (z)}  \frac {dk}{2 \pi^2} k^2 P_0(k,z) = \mbox{const.} \, .
\ee
Here we have set $k_{\min} = 10^{-3} $ $h$/Mpc, as this roughly corresponds to the effective volume of our Euclid-like survey. The obtained values of $k_{\max}(z)$
can then be translated into ${l}_{\max}$ by using the approximation
\be
{l}^{\epsilon}_{\max} \simeq k^{\epsilon}_{\max} \, r_K - 1/2 \, .
\ee
In this way the choice  $k_{\max} = 0.15 \, h/$Mpc at $z=0$ translates into imposing that the variance is fixed to $\sigma^2(z) = 0.36 $, i.e. the r.m.s. of the perturbations is $\sigma \simeq 0.6$, meaning that perturbation theory is generally respected (peaks of order $\delta \sim 1$ are rare), and so this is a conservative choice. To be even more conservative, we impose that in any case the smallest scale can not go above $k_{\max}^{\mathrm{abs}} = 0.3 $ $h/$Mpc, as was similarly done by \citet{Wang:2010gq}. In practice, this maximum $k$ is reached around $z \simeq 1$. We have checked that at this redshift, the non-linear contribution to the power spectrum does not exceed $\sim 13 \%$, and is decreasing at higher redshifts.

We can see the results for Euclid, local PNG in Fig. \ref{fig:fish_single}, and all the marginalised 1D error bars in Fig.~\ref{fig:all1D}. The results are also summarised in detail in  Tables~\ref{tab:bigtable}, \ref{tab:bigtable_ort}, \ref{tab:bigtable_equi} for the three PNG configurations.
We remind the reader that in this case the results are marginalised over nuisance parameters for the Gaussian part of galaxy bias, the Finger-of-God effect, and where indicated also shot noise.

\subsection {Galaxy clustering, 2D}

In addition to the 3D galaxy clustering in redshift space discussed in the previous section, we have also calculated forecasts for the projected clustering
on the celestial sphere. This is done for several reasons: first, in this way it is  much easier to  combine the results with weak lensing in a consistent way, as we will discuss in the next section. Secondly, this allows us to directly compare the performances of the photometric and spectroscopic surveys of the Euclid-like mission in terms of galaxy clustering: a priori, it is not obvious whether the
tighter constraints on cosmological parameters could come from the more numerous galaxies with photometric redshifts or from the less rich but more accurately
located spectroscopic sample.
 Thirdly, we can thus present forecasts for the galaxy clustering of the DES, which does not include, at least in its initial form, a spectroscopic survey.

The 2D galaxy spectrum between a pair of redshift bins $i,j$ is a projection of the 3D spectrum which, in analogy with the weak lensing case above, can be written as 
\ba \label{eq:clgg}
 C^{g_i g_j}_l &=& \frac{2}{\pi} \int_0^{\infty}  d \beta \, \beta^2 \, \int_0^{\infty} dr_1 \,W^{g_i} (\beta, r_1) \, \Phi_{{l}}^{\beta}(r_1) 
\nonumber \\ &~& \cdot \, \int_0^{\infty}
dr_2 \, W^{g_j}(\beta, r_2) \, \Phi_{{l}}^{\beta}(r_2)  \,  \cdot  \,  P_{m} (\beta, r_1, r_2) \, .
\ea
This can then be written in Limber approximation in analogy with Eq.~(\ref{eq:Limber}). The sources in a bin $i$ are now given by
\be \label {eq:Wg}
W^{g_i}(\beta, r) = \frac{1}{N_i} \,  \frac {dN_i}{dr} (r)  \, b_{g_i}[r, k(\beta)].
\ee
The observed 2D galaxy power spectra $\tilde C^{g_i g_j}_l$ can again be modelled as a sum of the theoretical spectra and the noise. For each pair of redshift bins $i,j$, we have
\be
  \tilde C^{g_i g_j}_l  = C^{g_i g_j}_l +  N^{g_i g_j}_l \, ,
\ee
where the noise term $N^{g_i g_j}_l$ is now due to the shot noise:
\ba
N_l^{g_i g_j}              &=&  \delta_{ij} \frac {1}{\bar n_{i}} \, .
\ea
The Fisher matrix in this case is defined exactly as in Eq.~(\ref{eq:fish2d}) above, with the lensing power spectra replaced by the galaxy-galaxy ones.

For completeness and for the purpose of comparison, we will use in this case both photometric and spectroscopic data sets for Euclid, as described in the above sections.
Note that, as already mentioned for the 3D case, since here we are not accounting for bias non-linearities nor scale-dependences at small scales, Eq.~(\ref{eq:clgg}) is exact only in the linear regime; therefore 
we use the same limit on the small scales described above for both cases, i.e.
 we  assume now $k_{\max} = 0.15 \, h/$Mpc at $z=0$, evolved in redshift as in the 3D case described above.
  We will discuss in Section~\ref{sec:smallscales} the effects of changing this limit. Also in this case, we use the Limber approximation for $l \ge 200$, and the exact calculation for larger scales, so that we can extend the analysis all the way to $l_{\min} = 5 $.
See Fig.~\ref{fig:Cls} for a comparison of the Limber power spectra with the exact calculations for the range of projected observables we consider. Here we can see how the Limber approximation departs from the exact calculation at large scales for the  galaxy spectra.

\begin {figure*} 
\begin{center}
\begin{minipage}[t]{0.49\linewidth}
\includegraphics[width=1.05\textwidth,angle=0]{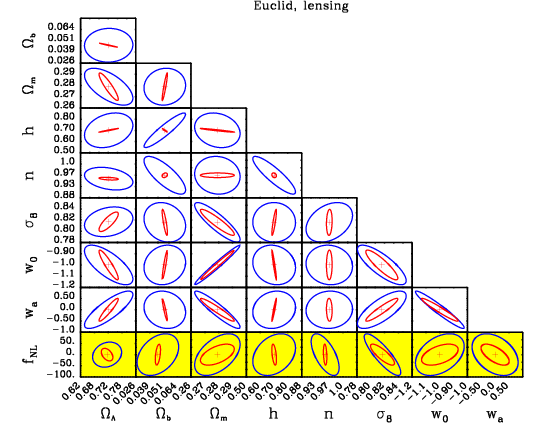}
\vspace{0.25cm}
\vfill
\includegraphics[width=1.05\textwidth,angle=0]{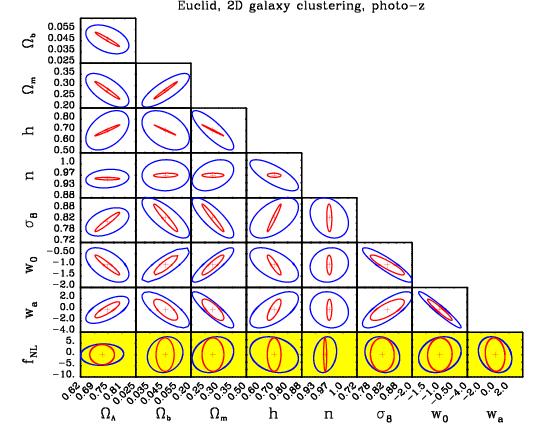}
\end{minipage}  \hfill
\begin{minipage}[t]{0.49\linewidth}
\includegraphics[width=1.05\textwidth,angle=0]{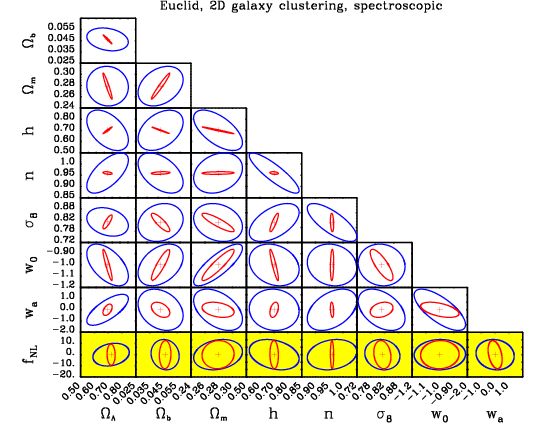}
\vspace{0.25cm}
\vfill
\includegraphics[width=1.05\textwidth,angle=0]{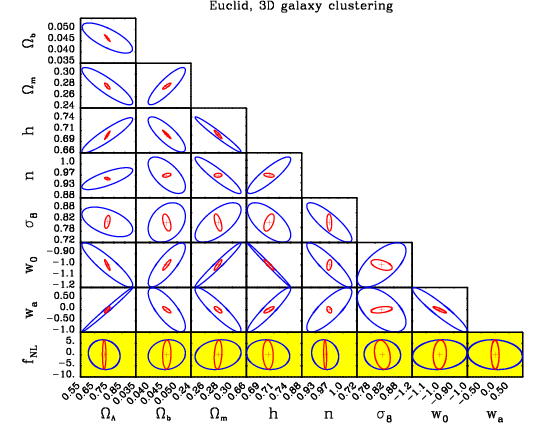}
\end{minipage}
\end{center}
\caption{Fisher matrix forecasts for the Euclid-like survey, using weak lensing (photometric survey), 2D galaxy clustering (photometric and spectroscopic surveys), and 3D galaxy clustering (spectroscopic only). For lensing, the used multipoles are from $l_{\min} = 5$ to $l_{\max}=20~000$, while for clustering the maximum mode is $k_{\max} = 0.15$ $h$/Mpc at $z=0$.  The forecasted posteriors are marginalised over the other not shown parameters. The blue ellipses refer to Euclid only, while in red we show the results including Planck priors. Notice that the axes ranges are different, which is necessary given the different constraining power of the different observables.}
\label{fig:fish_single}
\end{figure*}

We can see the results in Fig. \ref{fig:fish_single} for both photometric and spectroscopic parts of Euclid for the local PNG case, and all the marginalised 1D error bars in Fig.~\ref{fig:all1D}.  The results are also summarised in detail in  Tables~\ref{tab:bigtable}, \ref{tab:bigtable_ort}, \ref{tab:bigtable_equi} for the three PNG configurations.
These results are marginalised over nuisance parameters for the Gaussian part of galaxy bias.
\subsection {Combined results}

Finally, we shall combine the constraints from lensing and galaxy clustering in the projected case, as it is the only case where it is straightforward to do it consistently. Since the two probes are based on the same density field, it is crucial to correctly include their full covariance; on the other hand, it is also possible to include the lensing-clustering cross-correlation as a signal. These two operations can be achieved by considering one single Fisher matrix, as done by \citet{Hu:2003pt}, and defined by
\be \label {eq:fishfull}
F^{x}_{\alpha \beta} = f_{\mathrm{sky}} \sum_{l=l_{\min}}^{l_{\max}} \frac{(2 l + 1) }{2} \, \mathrm{Tr} \left[ \mathbf D^{x}_{l \alpha} \,  \left( \tilde \mathbf C^{x}_l \right)^{-1} \, \mathbf D^{x}_{l \beta} \,  \left( \tilde \mathbf C^{x}_l \right)^{-1} \right] \, .
\ee
In this case, the matrices $\tilde \mathbf C^{x}_l, \tilde \mathbf D^{x}_l$ will contain all the combinations of 2D spectra: $\tilde C_l^{\epsilon_i \epsilon_j}$, $\tilde C_l^{\epsilon_i g_j}$, $\tilde C_l^{g_i g_j}$. 
 Notice that the noise on the cross-spectra is zero, since we are assuming that shape and Poisson noise are uncorrelated: $\tilde C_l^{\epsilon_i g_j} =  C_l^{\epsilon_i g_j}$.

\begin {figure*} 
\begin{center}
\begin{minipage}[t]{0.49\linewidth}
\includegraphics[width=1.05\textwidth,angle=0]{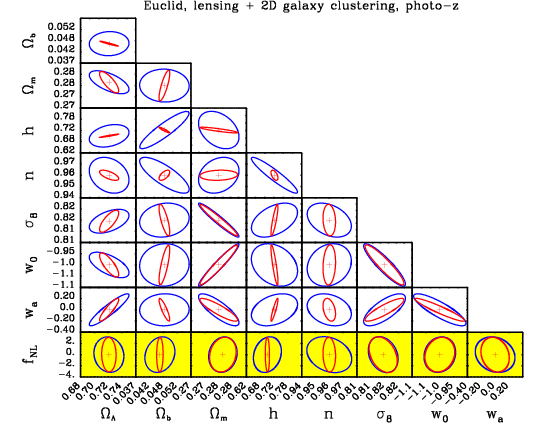}
\end{minipage}  \hfill
\begin{minipage}[t]{0.49\linewidth}
\includegraphics[width=1.05\textwidth,angle=0]{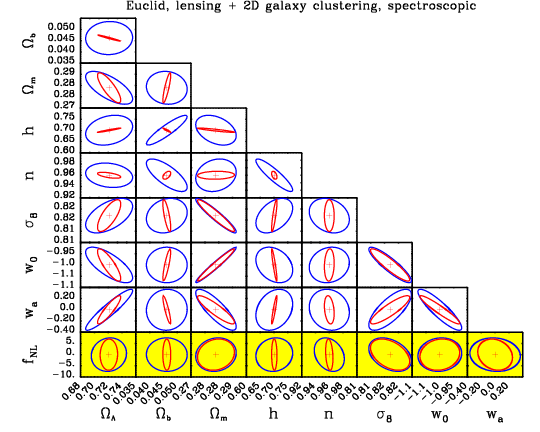}
\end{minipage}
\end{center}
\caption{Full combination of Euclid-like data. Fisher matrix forecasts for the combination of weak lensing (photometric survey) plus 2D galaxy clustering (photometric and spectroscopic surveys). Both the range of used multipoles  and the colour coding of the ellipses are as described above for the previous plots.}
\label{fig:fish_combine}
\end{figure*}

We consider in this case the combination of the lensing forecasts using the photometric catalogue with both photometric and spectroscopic galaxy clustering.
The combined results for Euclid are reported for both cases in Fig.~\ref{fig:fish_combine} for local PNG, and summarised in detail in  Tables~\ref{tab:bigtable}, \ref{tab:bigtable_ort}, \ref{tab:bigtable_equi} for the three PNG configurations. These results are marginalised over nuisance parameters for the Gaussian part of galaxy bias.

All the marginalised 1D error bars are shown in Fig.~\ref{fig:all1D} for the local PNG case.  We also present the results obtained for the DES specifications in Table~\ref{tab:bigtableDES} for the local case.
We can also see from Table~\ref{tab:testtabsum} the comparison between na\"ively summing the lensing and clustering Fisher matrices versus performing the full analysis. We can see here that the forecasted errors for the complete combination are smaller than the simplistic sum, in average by a factor of two. This points us to conclude that the effect of including all the cross-spectra between lensing and clustering as observable signals in the analysis is more important than the degrading effect which is produced by considering the covariances between the clustering and lensing auto-correlations.

\begin {table*}
\begin {center}
\begin{tabular}{l c c c c c c c c c}
\hline
  Euclid data   & $\sigma (\Omega_{\Lambda})$ & $\sigma (\Omega_b)$ & $\sigma (\Omega_m)$ & $\sigma (h)$ & $\sigma (n)$ & $\sigma (\sigma_8)$ & $\sigma (w_0)$ & $\sigma (w_a)$ & $\sigma (\fnl)$ \\
\hline
lensing &  .035  (.014)  &     .012  (.0016)  &     .011  (.0081)  &     .10  (.011)  &     .033  (.0040)  &     .017  (.011)  &     .12  (.087)  &     .55  (.34)  &     37  (19)  \\
2D clust. phot. &    .064  (.037)  &     .0085  (.0043)  &     .050  (.026)  &     .12  (.033)  &     .027  (.0036)  &     .042  (.028)  &     .51  (.29)  &     1.6  (.94)  &     3.2  (3.0)  \\
2D clust. spec. &     .087  (.019)  &     .0067  (.0027)  &     .023  (.015)  &     .12  (.020)  &     .059  (.0039)  &     .042  (.020)  &     .12  (.092)  &     .91  (.33)  &     6.7  (6.2)  \\
3D clust.  &  .063  (.0079)  &     .0041  (.00096)  &     .018  (.0036)  &     .020  (.006)  &     .024  (.0033)  &     .030  (.016)  &     .10  (.030)  &     .53  (.090)  &     4.4  (4.2)  \\
 ---  +shot noise  & .076  (.0082)  &     .0044  (.00098)  &     .022  (.0036)  &     .026  (.0062)  &     .032  (.0035)  &     .044  (.019)  &     .13  (.031)  &     .65  (.094)  &     4.4  (4.2)  \\
\hline
lens. + 2D phot.  &  .014  (.0069)  &     .0035  (.00077)  &     .0034  (.0032)  &     .033  (.0050)  &     .011  (.0030)  &     .0035  (.0032)  &     .037  (.035)  &     .19  (.13)  &    2.8  (2.6)  \\
lens. + 2D spec. & .017  (.0083)  &     .0049  (.00094)  &     .0048  (.0044)  &     .044  (.0064)  &     .014  (.0035)  &     .0051  (.0047)  &     .052  (.047)  &     .24  (.17)  &     4.9  (4.6)  \\

\hline
\end{tabular}
\caption{Expected marginalised errors from Euclid-like data for all parameters for the local non-Gaussianity case. The numbers
  within parentheses include the forecasted priors from Planck CMB
  temperature power spectrum. The additional line for the 3D clustering includes marginalisation over the shot noise parameters in each redshift bin.}
\label{tab:bigtable}
\end{center}
\end {table*}
%
%
%
%
%
\begin {table*}
\begin {center}
\begin{tabular}{l c c c c c c c c c}
\hline
 Euclid data   & $\sigma (\Omega_{\Lambda})$ & $\sigma (\Omega_b)$ & $\sigma (\Omega_m)$ & $\sigma (h)$ & $\sigma (n)$ & $\sigma (\sigma_8)$ & $\sigma (w_0)$ & $\sigma (w_a)$ & $\sigma (\fnl)$ \\
\hline
lensing &  .035  (.014)  &     .013  (.0016)  &     .011  (.0080)  &     .10 (.011)  &     .035  (.0040)  &     .017  (.011)  &     .12  (.087)  &     .55  (.33)  &     4.9  (2.4)  \\
2D clust. phot. &   .064  (.039)  &     .010  (.0046)  &     .051  (.028)  &     .12  (.035)  &     .040  (.0039)  &     .11  (.050)  &     .50  (.31)  &     1.6  (.96)  &     28  (8.1)  \\
2D clust. spec. &  .097  (.020)  &     .010  (.0029)  &     .023  (.016)  &     .15  (.021)  &     .086  (.004)  &     .17  (.045)  &     .12  (.096)  &     .95  (.32)  &     50  (10)  \\
3D clust. + shot  & .076  (.0083)  &     .0045  (.0010)  &     .022  (.0036)  &     .026  (.0062)  &     .032  (.0036)  &     .23  (.043)  &     .13  (.031)  &     .65  (.095)  &     56  (12)  \\
\hline
lens. + 2D phot.   &   .014  (.0076)  &     .0040  (.00088)  &     .0048  (.0039)  &     .034  (.0058)  &     .012  (.0033)  &     .0096  (.0064)  &     .049  (.041)  &     .24  (.19)  &     2.9  (1.6)  \\
lens. + 2D spec. &   .018  (.0087)  &     .0053  (.00099)  &     .0053  (.0048)  &     .047  (.0068)  &     .018  (.0036)  &     .010  (.0071)  &     .055  (.050)  &     .26  (.20)  &     3.4  (1.9)  \\
\hline
\end{tabular}
\caption{Expected marginalised errors from Euclid-like data for all parameters, for the orthogonal configuration. The constraints on $\fnl$ from the scale-dependent bias are degraded, while the results from lensing are in this case stronger. 
}
\label{tab:bigtable_ort}
\end{center}
\end {table*}
%

%
\begin {table*}
\begin {center}
\begin{tabular}{l c c c c c c c c c}
\hline
Euclid data   & $\sigma (\Omega_{\Lambda})$ & $\sigma (\Omega_b)$ & $\sigma (\Omega_m)$ & $\sigma (h)$ & $\sigma (n)$ & $\sigma (\sigma_8)$ & $\sigma (w_0)$ & $\sigma (w_a)$ & $\sigma (\fnl)$ \\
\hline
lensing &     .035  (.014)  &     .012  (.0016)  &     .011  (.0082)  &     .10  (.011)  &     .031  (.0040)  &     .017  (.012)  &     .12  (.088)  &     .54  (.34)  &     17  (9.2)  \\
2D clust. phot. &  .085  (.041)  &     .0098  (.0047)  &     .057  (.029)  &     .12  (.036)  &     .038  (.0039)  &     .12  (.057)  &     .51  (.32)  &     1.7  (1.0)  &     100  (30)  \\
2D clust. spec. &     .11  (.020)  &     .0069  (.0028)  &     .024  (.016)  &     .12  (.020)  &     .073  (.0040)  &     .16  (.048)  &     .14  (.094)  &     1.1  (.31)  &     150  (35)  \\
3D clust. + shot  &   .079  (.0083)  &     .0045  (.00099)  &     .024  (.0036)  &     .027  (.0062)  &     .038  (.0036)  &     .25  (.043)  &     .14  (.031)  &     .69  (.094)  &     220  (37)  \\
\hline
lens. + 2D phot.   &     .014  (.0076)  &     .0040  (.00088)  &     .0052  (.0039)  &     .033  (.0058)  &     .013  (.0033)  &     .011  (.0066)  &     .050  (.041)  &     .24  (.18)  &     12  (6.5)  \\
lens. + 2D spec. &   .017  (.0090)  &     .0050  (.0010)  &     .0058  (.0049)  &     .044  (.0071)  &     .015  (.0035)  &     .011  (.0076)  &     .057  (.051)  &     .28  (.21)  &     11  (7.0)  \\
\hline
\end{tabular}
\caption{Expected marginalised errors from Euclid-like data for all parameters for the equilateral configuration. The bias is kept scale-independent here.
 The constraints on $\fnl$ from galaxy clustering are strongly weakened, as the bias is now scale-independent. The results from lensing remain instead  strong.
}
\label{tab:bigtable_equi}
\end{center}
\end {table*}

\begin {table*}
\begin {center}
\begin{tabular}{l c c c c c c c c c}
\hline
DES data &  $\sigma (\Omega_{\Lambda})$ & $\sigma (\Omega_b)$ & $\sigma (\Omega_m)$ & $\sigma (h)$ & $\sigma (n)$ & $\sigma (\sigma_8)$ & $\sigma (w_0)$ & $\sigma (w_a)$ & $\sigma (\fnl)$ \\
\hline
lensing &  .20  (.033)  &     .036  (.0036)  &     .037  (.022)  &     .32  (.027)  &     .13  (.0042)  &     .063  (.030)  &     .41  (.26)  &     2.6  (.96)  &     150  (46)  \\
2D clust. phot. &  .23  (.050)  &     .019  (.0056)  &     .10  (.033)  &     .34  (.042)  &     .10  (.0041)  &     .058  (.039)  &     .82  (.44)  &     3.3  (2.0)  &     12  (11)  \\
\hline
lens. + 2D phot. &   .062  (.013)  &     .014  (.0014)  &     .0082  (.0074)  &     .12  (.010)  &     .039  (.0041)  &     .0094  (.0086)  &     .093  (.090)  &     .61  (.35)  &     8.6  (8.2)  \\
\hline
simple sum &  .12  (.028)  &     .014  (.0030)  &     .024  (.018)  &     .13  (.022)  &     .040  (.0041)  &     .023  (.018)  &     .28  (.22)  &     1.6  (.77)  &     11  (10)  \\
\hline
\end{tabular}
\caption{Expected marginalised errors for all parameters for DES for the local PNG case. The numbers within parentheses include the forecasted priors from Planck CMB temperature power spectrum.}
\label{tab:bigtableDES}
\end{center}
\end {table*}

\begin {table*}
\begin {center}
\begin{tabular}{l c c c c c c c c c}
\hline
Euclid  data   & $\sigma (\Omega_{\Lambda})$ & $\sigma (\Omega_b)$ & $\sigma (\Omega_m)$ & $\sigma (h)$ & $\sigma (n)$ & $\sigma (\sigma_8)$ & $\sigma (w_0)$ & $\sigma (w_a)$ & $\sigma (\fnl)$ \\
\hline
lensing  &     .035  (.014)  &     .012  (.0016)  &     .011  (.0081)  &     .10  (.011)  &     .033  (.0040)  &     .017  (.011)  &     .12  (.087)  &     .55  (.34)  &     37  (19)  \\
2D clust. phot. &   .064  (.037)  &     .0085  (.0043)  &     .050  (.026)  &     .12  (.033)  &     .027  (.0036)  &     .042  (.028)  &     .51  (.29)  &     1.6  (.94)  &     3.2  (3.0)  \\
\hline
lens. + 2D phot.  &  .014  (.0069)  &     .0035  (.00077)  &     .0034  (.0032)  &     .033  (.0050)  &     .011  (.0030)  &     .0035  (.0032)  &     .037  (.035)  &     .19  (.13)  &    2.8  (2.6)  \\
\hline
simple sum  &  .024  (.011)  &     .0040  (.0012)  &     .0079  (.0066)  &     .038  (.0089)  &     .011  (.0031)  &     .0075  (.0065)  &     .089  (.074)  &     .39  (.25)  &     2.8  (2.7)  \\
\hline
\end{tabular}
\caption{Comparison of summing two matrices obtained for Euclid versus combining them appropriately including all the cross-signals. Results between parentheses include Planck priors.}
\label{tab:testtabsum}
\end{center}
\end {table*}

\section {Discussion and additional analyses} \label{sec:additional}

We present here some additional results which extend our basic analysis of the previous section, and discuss some important complementary issues.

\subsection {Additional priors}
We first study the effect of adding as a prior the forecasted constraints for the Planck CMB satellite, whose results will be available by the time these surveys are completed. The Fisher matrix for Planck is obtained in analogy with the galaxy 2D case, and the CMB spectra are calculated with the CAMB CMB code~\citep{Lewis:1999bs}. We describe this in more detail in the Appendix~\ref{app:planck}.\footnote{We acknowledge Jochen Weller for supplying us with the CMB Fisher matrix code.}

 The reduced likelihood intervals are also shown in Figs.~\ref{fig:fish_single},~\ref{fig:fish_combine},~\ref{fig:all1D}.
We can see that in this case all the constraints improve greatly, with the significant exception of $\fnl$. This is because the Planck prior we are using is only taking into account the CMB power spectrum, and not the bispectrum, which is where most information on non-Gaussianity is. In other words, our Planck Fisher matrix does not depend on $\fnl$ at all.

\subsection {Summary of the results}
\begin {figure*} 
\begin{center}
\includegraphics[width=0.8\textwidth,angle=0]{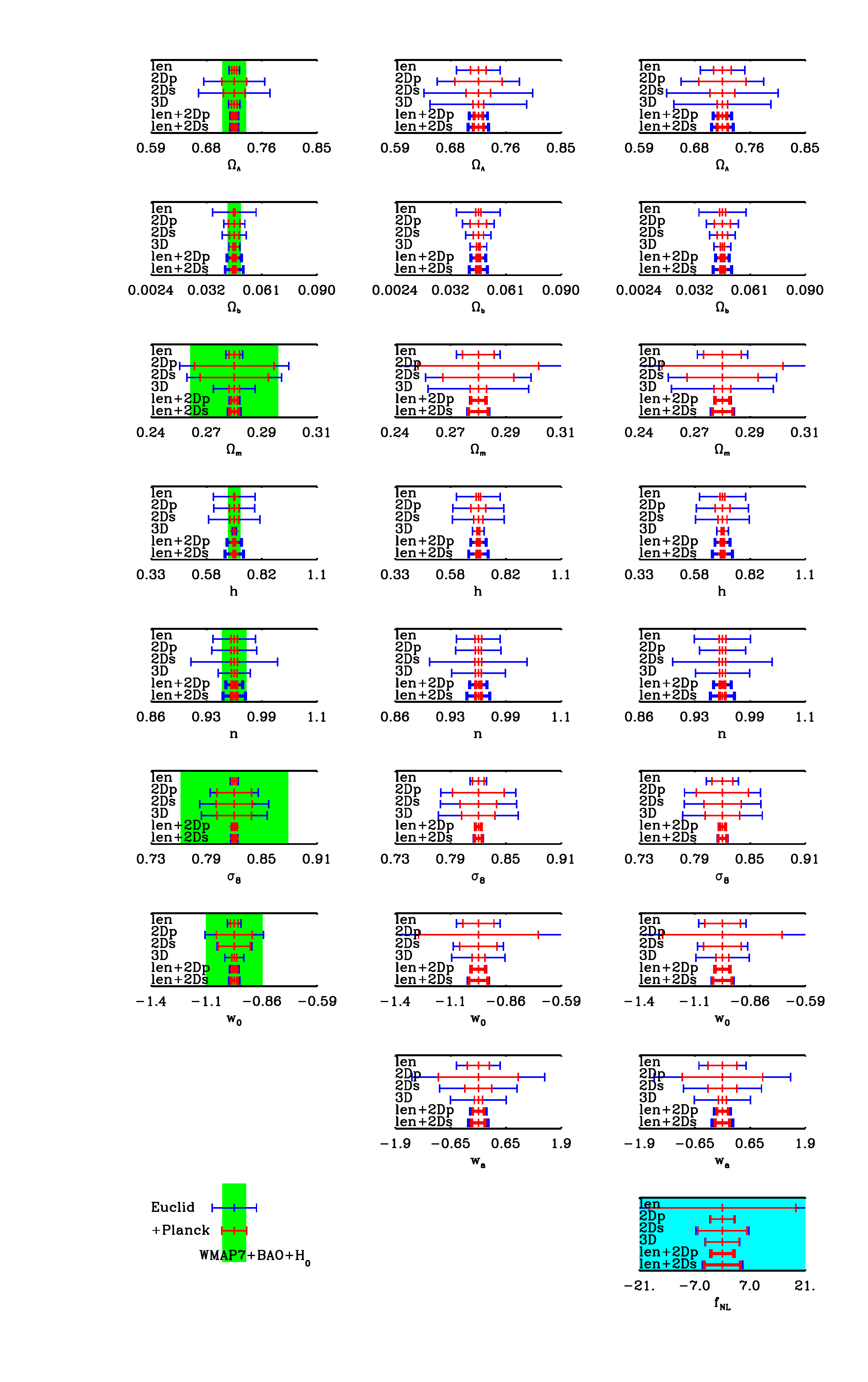} 
\end{center}
\caption{Marginalised 1D errors from Euclid-like data on each cosmological parameter, for a wCDM model (left), and with the addition of $w_a$ (centre) and $\fnl$ (right), for the local configuration. For each parameter, we show the result obtained using all four observables and their combinations. The error bars in red include the Planck priors. We also overlay the current constraints from WMAP7, baryon acoustic oscillations and $H_0$ measurements in light green, and the size of the current uncertainty on $\fnl$ from the CMB bispectrum in cyan. The results on the standard parameters remain unaffected by the addition of $\fnl$.}
\label{fig:all1D}
\end{figure*}

We summarise in Fig.~\ref{fig:all1D} all the marginalised 1D constraints which are obtained from the Fisher matrix analysis in the Gaussian and non-Gaussian cases, with and without the imposition of the Planck priors. Here we can also see a comparison between the forecasted future results for Euclid and the current constraints from the WMAP satellite and other probes. It is particularly interesting to notice that the errors on the standard cosmological parameters remain largely unchanged when $\fnl$ is added. We also found by looking at the covariant elements in the Fisher matrices that $\fnl$ is almost completely uncorrelated from the other parameters, which is due to the very particular scale-dependent behaviour which it produces on the observables.

\subsection {The different shapes of non-Gaussianity} \label{sec:equi}

We now focus on the differences between the three types of primordial non-Gaussianity described in Section~\ref{sec:NG}: namely the local, equilateral and 
orthogonal bispectrum shapes.
In all these cases PNG alters both the matter power spectrum through the 
halo mass function (which depends on the skewness of the linear density field), and the galaxy bias.

The different forms for the skewness are described in the Appendix~\ref{sec:appbias}, while we use the expressions for the corrections to the bias in Eqs.~(\ref{eq:dball}) derived by \citet{Schmidt:2010gw} and improved by \citet{2011PhRvD..84f1301D}.
These expressions are approximated but are rather accurate at small wavenumbers 
for the local and orthogonal cases \citep{Schmidt:2010gw} where the scale-dependent part of the bias scales as $k^{-2}$ and $k^{-1}$, respectively.
On the other hand, in the equilateral case, the bias becomes asymptotically constant
for small $k$ and the only scale-dependence of the bias comes from the
linear transfer function appearing in the function $\alpha$.
Unfortunately, the numerical factor $ \frac{\sigma_{R, -2}^2}{\sigma_{R,0}^2}$ which appears in the bias correction is typically large, so that
the scale-dependent deviations sourced by $T(k)$ easily  outweigh the effect 
of the mass-density skewness in the matter power spectrum at large $k$.
This is not ideal, as Eqs.~(\ref{eq:dball}) have been derived with the peak-background split technique, which should only hold on large
scales.
For this reasons we decide here to be conservative, and to keep the bias 
fully scale-independent in the equilateral case, fixing it at all scales to its small-$k$ value.

The marginalised constraints on the non-Gaussianity parameter $\fnl$ are shown in Table~\ref{tab:equil} for the three configurations.
Here we can see that the constraints in the orthogonal and equilateral cases get generally weaker for the clustering probes due to the reduced importance of the scale-dependent bias. The constraints from lensing remain nevertheless quite strong, as the effect of the skewness is independent from the bias; the results for the combined probes are therefore very promising for all the configurations. 

As a further note of caution, we have nevertheless to stress that, differently from the local case, we could not check with N-body simulations the predictions of the non-linear regime for the orthogonal and equilateral cases. This means that the constraints from lensing and its combinations in these two cases are necessarily less reliable than both the local case and the constraints from the scale-dependent bias, whose predictions are well tested and understood.

Finally, we can compare Tables~\ref{tab:bigtable}, \ref{tab:bigtable_ort}, \ref{tab:bigtable_equi} to see how much the errors on the other parameters change. Here we can see that when the constraint on $\fnl$ degrades greatly, such as in the galaxy clustering probes alone, the constraints on some other parameters also degrade significantly, especially $\sigma_8$. However the combined lensing+clustering result stays largely unchanged.

Our constraints on $\fnl$ are weaker than what was found by \citet{Fedeli:2010ud} for the 3D case only, as these authors calculated forecasts by varying only $\sigma_8$ and $\fnl$ while fixing the rest of the parameters to \LCDM and also used a more optimistic cutoff than ours for the non-linear regime.

\begin {table}
\begin {center}
\begin{tabular}{l c c c} 
\hline
Euclid data                                &  \multicolumn{3}{c}{ $\sigma (\fnl)$ }    \\
    (with Planck priors)       &   local   &   orthogonal   &   equilateral \\ 
\hline
lensing                        &  37 (19)  & 4.9 (2.4) & 17 (9.2) \\ 
2D clust., photo-z             & 3.2 (3.0) &  28 (8.1) & 100 (30) \\
2D clust., spectro             & 6.7 (6.2) &  50 (10) & 150 (35)\\
3D clust. + shot              & 4.4 (4.2) &  56 (12) & 220 (37) \\ 
\hline
lens. + 2D photo               & 2.8 (2.6) & 2.9 (1.6) & 12 (6.5)\\ 
lens. + 2D spectro             & 4.9 (4.6) & 3.4 (1.9) & 11 (7.0) \\ 
\hline
\end{tabular}
\caption{Forecasted results for non-Gaussianity for the different configurations. Constraints from galaxy clustering weaken in the orthogonal and equilateral cases, as the scale dependence of the bias is reduced. For the lensing (and thus also combined) cases, the constraints remain strong, as the non-Gaussian mass function is in all configurations expected to deviate at similar levels from the Gaussian model. } 
\label{tab:equil}
\end{center}
\end {table}

\subsection {Dependence on the fiducial}

We then study the dependence of our forecasts on the choice of the fiducial values for the parameters, in particular extending the analysis to
 $\fnl \ne 0$ (in the case of local PNG only), or $w_0 \ne -1$. For this purpose, we run our full analysis for two non-standard cases, where the fiducials have the value of one parameter altered to $w_0 = -0.95 $ and $\fnl = 30$ respectively.

\subsubsection*{Fiducial with dark energy}

In the case of dark energy, we calculated the forecasted errors changing the value of the fiducial equation of state from $w_0 = -1$ to $w_0 = -0.95$. We obtain that the constraints do not change significantly, as we can see in Table~\ref{tab:w0fid}; we can conclude that the model with dark energy of $w_0 = -0.95$ would be effectively detected by Euclid, even without the need of Planck CMB priors.

\begin {table}
\begin {center}
\begin{tabular}{l c c}
\hline
Euclid data                                &  \multicolumn{2}{c}{ $\sigma (w_0)$ }    \\
    (including Planck priors)       &   $w_0=-1.0$   &   $w_0=-0.95$ \\
\hline
lensing                          & .12 (.087)   & .11  (.079)  \\
2D clustering, photo-z           & .51 (.29)    & .49 (.29)  \\
2D clustering, spectroscopic     & .12 (.092)   & .091 (.084) \\
3D clust + shot noise                & .13 (.031)   &  .12 (.031)    \\
\hline
lensing + 2D photo               & .037 (.035) &  .035 (.033) \\
lensing + 2D spectro            &  .052 (.047) & .040 (.039) \\
\hline
\end{tabular}
\caption{Forecasted results for non-Gaussianity as a function of the
  fiducial value of $w_0$, without and with Planck priors. These results show that a Euclid-like survey should be able to distinguish between a cosmological constant and a different dark energy model.} 
\label{tab:w0fid}
\end{center}
\end {table}

\subsubsection* {Non-Gaussian fiducial}

We chose to look at a local non-Gaussian model with $\fnl = 30$, as this is near the peak of the current posterior probability distribution given current data.
We found that setting the fiducial to such a model degrades to some extent the accuracy of the errors. For the combined case (lensing + 2D photometric galaxy clustering), the uncertainty on $\fnl$ degrades from $\sigma(\fnl) = 2.6$ to $\sigma(\fnl) =4.1$ when Planck priors are included. We can see the results summarised in Table~\ref{tab:fnlfid}, from which we can conclude that this model would be detected with high significance by Euclid.

\begin {table}
\begin {center}
\begin{tabular}{l c c}
\hline
Euclid data                                &  \multicolumn{2}{c}{ $\sigma (\fnl)$ }    \\
    (including Planck priors)       &   $\fnl=0$   &   $\fnl=30$ \\
\hline
lensing                          & 37 (19)  &    38 (18) \\
2D clustering, photo-z           & 3.2 (3.0)  & 4.9 (4.3) \\
2D clustering, spectroscopic     & 6.7 (6.2) &  8.1 (7.3) \\
3D clust + shot noise                 & 4.4 (4.2)  & 5.0 (4.7)   \\
\hline
lensing + 2D photo            &  2.8 (2.6) &  5.1  (4.1)  \\
lensing + 2D spectro          &  4.9 (4.6) & 5.4 (5.0) \\
\hline
\end{tabular}
\caption{Forecasted results for non-Gaussianity as a function of the
  fiducial value of $\fnl$ without and with Planck priors: detection versus upper limit scenarios. These results show that a Euclid-like survey will be able to either detect models with  primordial non-Gaussianity down to the level of $\fnl \sim 10$, or to find strict constraints around a Gaussian model.}
\label{tab:fnlfid}
\end{center}
\end {table}

\subsubsection*{Galaxy biasing} \label{sec:bchange}
The scale-dependent non-Gaussian correction to the galaxy linear bias is proportional to the Lagrangian bias $b_g-1$. Therefore, in the limit where the scale-independent part is $b_g(z) \equiv 1$, 
the effect of PNG will vanish making it impossible to constrain $\fnl$ from
galaxy clustering on large scales.
On the other hand,
it is interesting to notice that in the presence of highly biased tracers, such as luminous red galaxies (LRG) or quasars, the effect -- and the constraints -- are maximised.

We  look at how much the constraints are modified if we alter our fiducial bias by a fraction $f_b$, i.e. we take 
\be
b_g^{\mathrm{fid}}(z) = 1 + f_b \, \left(\sqrt{1+z} - 1 \,\, \right) \, ,
\ee
and we consider two cases with $f_b = 1 \pm 0.25$. We can see in Table~\ref{tab:fnlfb} that the constraints on $\fnl$ are indeed strongly dependent on this choice, as expected, except obviously for the lensing case where there is no biasing.

\begin {table}
\begin {center}
\begin{tabular}{l c c c}
\hline
Euclid data                                &  \multicolumn{3}{c}{ $\sigma (\fnl)$ }    \\
    (including Planck priors)       & $f_b = 0.75 $ &  $f_b = 1 $   &   $ f_b = 1.25 $ \\
\hline
lensing                             &  37 (19)    & 37 (19)    &  37 (19)   \\
2D clustering, photo-z              &  3.8 (3.6)   &  3.2 (3.0) & 2.8 (2.6) \\
2D clustering, spectroscopic        &  8.6 (7.9)  &  6.7 (6.2) & 5.8 (5.6) \\
3D clust + shot noise                    & 5.7 (5.5)   & 4.4 (4.2)  &  3.6 (3.5) \\
\hline
lensing + 2D photo                  & 4.2 (3.7)  & 2.8 (2.6)  &  2.4  (2.3)  \\
lensing + 2D spectro                & 6.3 (5.7)  & 4.9 (4.6)  &  4.1 (3.9) \\
\hline
\end{tabular}
\caption{Marginalised forecasted results from Euclid for non-Gaussianity as a function of the
  fiducial value of the bias $b_g^{\mathrm{fid}}(z)$, without and with Planck priors.}
\label{tab:fnlfb}
\end{center}
\end {table}

\subsection {Scale-dependent non-Gaussianity} \label{sec:nfnl}

The local shape of non-Gaussianity
can be obtained e.g. when multiple scalar fields give a contribution to
the curvature perturbations. However,
in many models of inflation, non-Gaussianity is generated in a scale-dependent way. For this reason, a new parameter has been introduced by \citet{Chen:2005fe} in the equilateral case and by \citet{Byrnes:2009pe} in the local case: the spectral index of non-Gaussianity, $n_{\fnl}$, defined so that the effective $\fnl(k)$ is written as
\be
\fnl (k) = \bar \fnl \, \left( \frac{k}{k_{\mathrm{piv}}} \right)^{n_{\fnl}} \, ,
\ee
where $\bar \fnl$ is the value of $\fnl$ at the pivot scale $k_{\mathrm{piv}}$.
We choose the pivot scale to be near the logarithmic centre of the expected data \citep{Cortes:2007ak} and fix it to $k_{\mathrm{piv}} = 0.02 $ $h/$Mpc.

For the simplest case of local non-Gaussianity, in which a single field is responsible for generating the curvature perturbation such as the curvaton scenario \citep{Huang:2010cy,Byrnes:2011gh},  it is possible to take the scale dependence of $\fnl$ out of the integrals, and simply apply it to the bias variation $\Delta b (k) $ \citep{Shandera:2010ei, 2011PhRvD..84f1301D}.  
In this case, by adding $n_{\fnl}$ to our parameter array, and fixing the fiducial $\bar \fnl = 30 $, we obtain the results of Table~\ref{tab:nfnl}. The best constraint comes from  combining lensing, clustering, and the Planck prior, giving $\sigma (n_{\fnl}) \simeq 0.12$. The constraints on $\fnl$ get in this case weakened.
These constraints are similar to to the forecasted limits from the Planck CMB bispectrum \citep{Sefusatti:2009xu}, and fully independent.

\begin {table}
\begin {center}
\begin{tabular}{l c c}
\hline
Euclid data       & $\sigma (\bar \fnl)$   &    $\sigma (n_{\fnl})$ \\
\hline
lensing                          & 68 (58)  &  .66 (.59) \\
2D clustering, photo-z         & 14  (9.6)   & .38 (.26) \\
2D clustering, spectroscopic   &  23  (14)  &  .64 (.38)   \\
3D clust + shot noise                 &  10 (7.6)  &  .28  (.21) \\
\hline
lensing + 2D photo             &  5.3 (4.3)  &  .18 (.14)   \\
lensing + 2D spectro          &  6.6 (5.2) & .17 (.12) \\
\hline
\end{tabular}
\caption{Forecasted results from Euclid-like data for scale-dependent non-Gaussianity and
  degraded constraints on the scale-independent part $\bar \fnl$. The fiducial model has here $\bar \fnl = 30$.}
\label{tab:nfnl}
\end{center}
\end {table}

\subsection {Smooth bias parametrisation}

\begin {table*}
\begin {center}
\begin{tabular}{l c c c c c c c c c}
\hline
Euclid data   & $\sigma (\Omega_{\Lambda})$ & $\sigma (\Omega_b)$ & $\sigma (\Omega_m)$ & $\sigma (h)$ & $\sigma (n)$ & $\sigma (\sigma_8)$ & $\sigma (w_0)$ & $\sigma (w_a)$ & $\sigma (\fnl)$ \\
\hline
2D clust. phot.   &   .054  (.029)  &     .0063  (.0037)  &     .032  (.022)  &     .099  (.028)  &     .027  (.0036)  &     .030  (.024)  &     .27  (.22)  &     .80  (.67)  &     3.0  (3.0)  \\
2D clust. spec.  &    .076  (.013)  &     .0062  (.0018)  &     .013  (.0093)  &     .089  (.013)  &     .047  (.0039)  &     .031  (.014)  &     .085  (.052)  &     .67  (.28)  &     7.0  (6.6)  \\
3D clust + shot  &   .078  (.0080)  &     .0043  (.00091)  &     .019  (.0030)  &     .026  (.0056)  &     .020  (.0036)  &     .0060  (.0042)  &     .13  (.031)  &     .67  (.10)  &     4.4  (4.1)  \\
\hline
lens. + 2D phot.    &   .014  (.0068)  &     .0035  (.00075)  &     .0033  (.0031)  &     .033  (.0049)  &     .011  (.0030)  &     .0034  (.0032)  &     .036  (.035)  &     .18  (.13)  &     2.8  (2.6)  \\
lens. + 2D spec.  &  .016  (.0071)  &     .0049  (.00081)  &     .0038  (.0034)  &     .043  (.0053)  &     .014  (.0035)  &     .0041  (.0038)  &     .039  (.035)  &     .21  (.13)  &     4.9  (4.6)  \\

\hline
\end{tabular}
\caption{Expected marginalised errors for Euclid-like data for all parameters, using a parametric form of the galaxy bias. The numbers
  within parentheses include the forecasted priors from Planck CMB
  temperature power spectrum. For 3D, includes marginalisation over FoG. }
\label{tab:bigtableparab}
\end{center}
\end {table*}
%

The constraints obtained so far are very conservative in the treatment of the bias. By assigning to every redshift bin an independent nuisance bias parameter, we are actually allowing for much more freedom than it is physically reasonable to expect. In particular, we have good reasons to expect the galaxy bias to be a smooth function of redshift. Therefore, we propose an alternative, less conservative, parametrisation, in which the bias is assumed to be a polynomial function of redshift. In the following, we will use a third-order polynomial:
\be \label{eq:polyb}
b(z) = b_0 + b_1 (z-1) + b_2 (z-1)^2 + b_3 (z-1)^3 \, ,
\ee
with fiducial values $b_0=\sqrt{2}, b_1 = (2 \sqrt 2)^{-1}, b_2 = -(16 \sqrt 2)^{-1}, b_3 = (64 \sqrt 2)^{-1}$. 
The expansion in $(z-1)$ is required if we want the bias to approximate $\sqrt{1+z}$ over the interval $0<z<2$. Notice that this expansion converges only in this interval, and thus it would have to be modified for a broader redshift range. We have also checked that by expanding up to the third order we are able to approximate the $\sqrt{1+z}$ function to better than 1\% in this range.

When including the shot noise marginalisation in this case, we also parametrise it as a third-order polynomial instead of leaving it as a completely free set of independent nuisance parameters for each bin, for the same reasons of requiring that physical quantities should have a smooth evolution in redshift.
The forecasted errors shrink in this case however only marginally, as we can see in Table~\ref{tab:bigtableparab}.

Finally, we evaluated how well could galaxy bias be measured with these observables. Referring to the smooth polynomial model of Eq.~(\ref{eq:polyb}), we calculated the marginalised errors on its coefficients, which can be seen for the different Euclid probes in Table~\ref{tab:bi}. Again, the combination of clustering+lensing gives the strictest results, since in this case all the other parameters are much better constrained, yielding an accuracy on the bias at the level of $1\%$.

\begin {table}
\begin {center}
\begin{tabular}{l c c c c}
\hline
Euclid data       & $\sigma (b_0)$   &    $\sigma (b_1)$  &    $\sigma (b_2)$  &    $\sigma (b_3)$ \\
\hline
2D clustering, photo-z         & .038  & .031  & .016  & .081    \\
2D clustering, spectro         & .067  & .038 & .017  & .026   \\
3D clust. + shot noise              & .015  &  .015   & .011  &  .012  \\
\hline
lensing + 2D photo             &  .0048  & .0047  & .0046  & .0045    \\
lensing + 2D spectro           &  .0066 &  .0076  & .012  & .024     \\
\hline
\end{tabular}
\caption{Forecasted results from Euclid for the measurement of the galactic bias. A third-order polynomial expansion is assumed.}
\label{tab:bi}
\end{center}
\end {table}

\subsection{Dependence on the cutoff scales} \label{sec:smallscales}

\subsubsection*{Small scales}
\begin{figure} 
\begin{center}
\includegraphics[width=0.5\textwidth,angle=0]{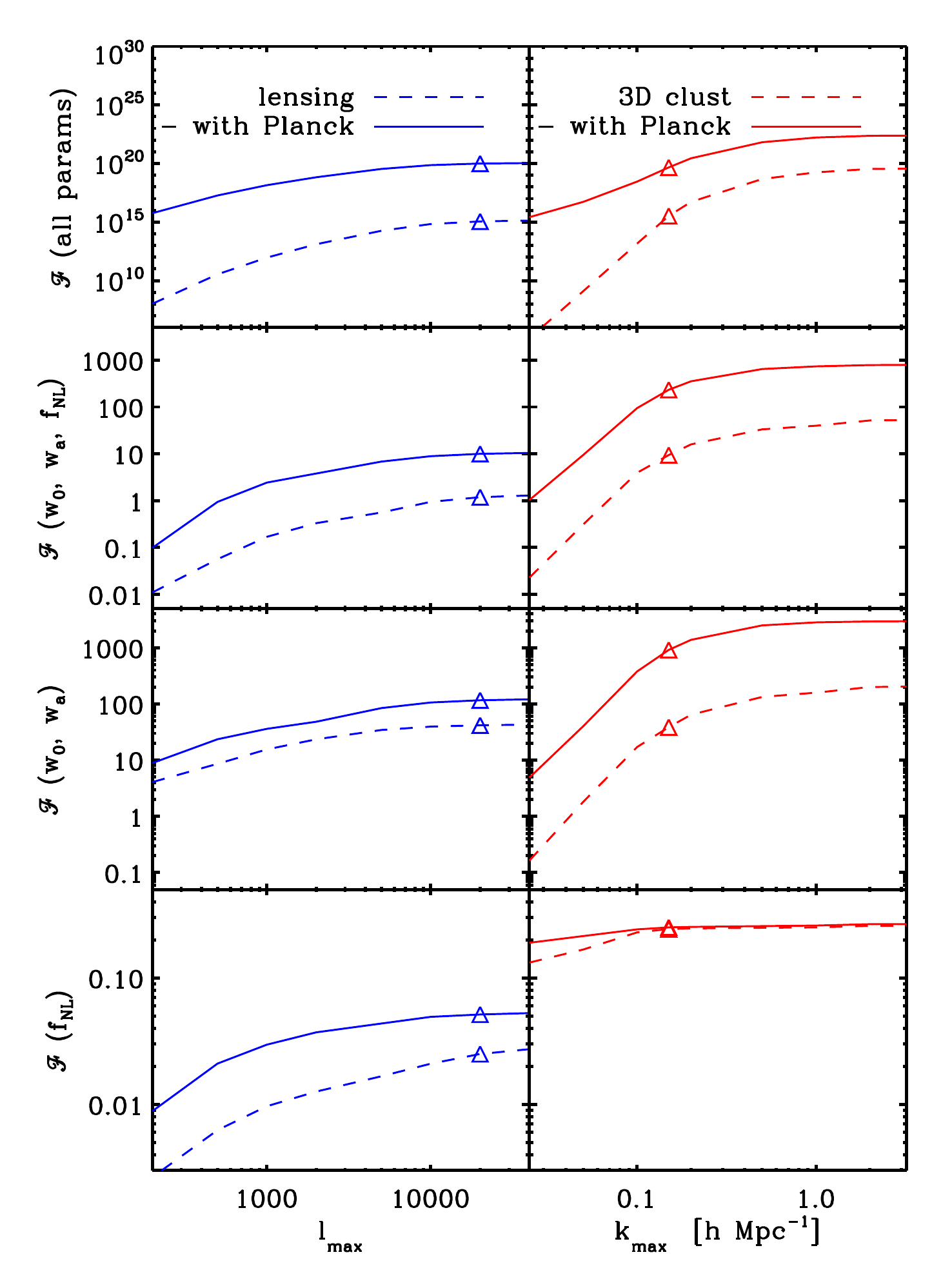}
\end{center}
\caption{Dependence of the Euclid-like constraints on the small-scale cuts. In the top panels we show the evolution of the full figure of merit as a function of the minimum scale considered in the analysis: $l_{\max}$ for the 2D part (left) and $k_{\max}$ at $z=0$ for the 3D part (right). In the following rows, we show the corresponding FoM relative to different combinations of few selected parameters: $w_0, w_a, \fnl$. The dashed lines denote Euclid only, while the solid lines include the Planck CMB priors (temperature power spectrum only). The triangles mark the scale which was chosen in the main analysis. 
}
\label{fig:lmax}
\end{figure}

Here we study how the constraints change if we vary the maximum mode $k_{\max}, l_{\max}$ used in the forecasts. Since the constraints on each individual parameter will not in general vary monotonically, we will study the most interesting quantity, the figure of merit (FoM) $\mathcal{F}$ of the survey with respect to a set of $N$ parameters $\boldsymbol{\vartheta}$, defined similarly to \citet{Albrecht:2006fa} as
\be
\mathcal{F}(\boldsymbol{\vartheta}) = \frac{1} {\prod_{i=1}^N \tilde \sigma (\vartheta_i) }  \, ,
\ee
where each $\tilde \sigma (\vartheta_i) $ is the width of the error ellipsoid along the axis defined by the i-th eigenvector of the Fisher matrix. In other words, it is easy to obtain their products by taking the determinant of the Fisher matrix as
\be
\mathcal{F}(\boldsymbol{\vartheta}) =  \sqrt{ \left| \mathbf{F}(\boldsymbol{\vartheta}) \right| } \,\, . 
\ee
We then study the total FoM,  the FoM obtained for the dark energy parameters only, and for the dark energy + $\fnl$ parameters.
We can see the result in Fig.~\ref{fig:lmax}, where we report the evolution of the FoM
for lensing and the 3D power spectrum as a function of  $l_{\max}$ and $k_{\max}$ respectively. In the 3D case, we evolve the scale cutoff with redshift as explained in the previous sections, by imposing that the variance of the density field stays constant in all redshift bins. In addition, to avoid considering exceedingly small scales, we impose a further condition that in every bin we will only consider scales $k \le k_{\max}^{\mathrm{abs}} = 2 k_{\max} (z=0)$.

 As expected, the results are strongly dependent on the choice of the minimum scales used. For the case of galaxy clustering, we can see that we could still largely improve the constraints by extending the analysis deeper into the non-linear regime, with the use of more advanced biasing models reliable on smaller scales. On the other hand we see that for weak lensing we have already included most of the signal with our choice of $l_{\max}$. It can be seen that more conservative choices would still yield a very similar result; for example by choosing more conservatively $l_{\max} = 5000$, we have that the marginalised error on $\fnl$ with Planck priors degrade from 19 to 22 only, and the total FoM degrades by only a factor of 3.

\subsubsection*{Large scales}
\begin{figure} 
\begin{center}
\includegraphics[width=0.5\textwidth,angle=0]{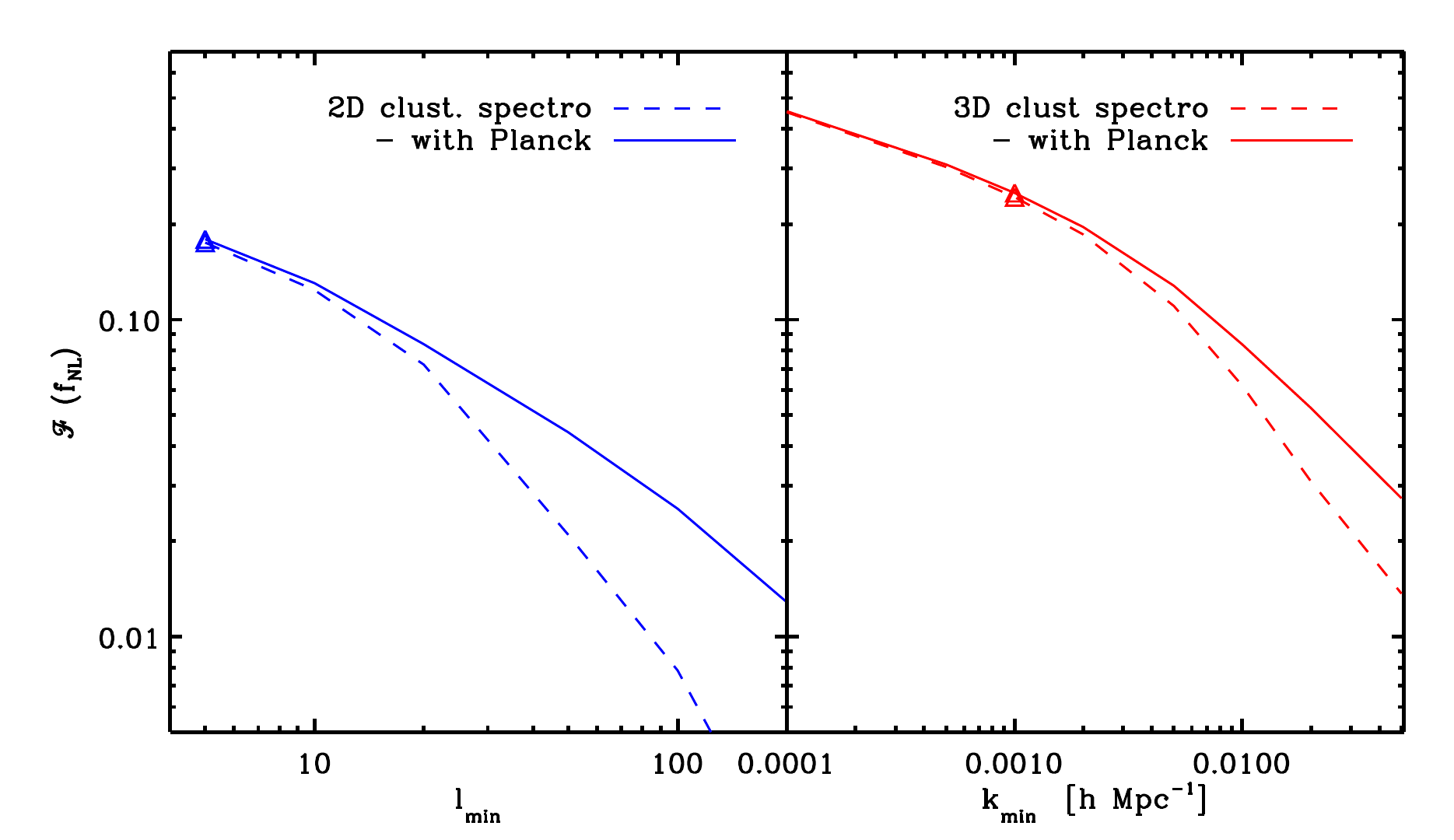}
\end{center}
\caption{Dependence of the Euclid constraints on the large-scale cuts for the 2D and 3D cases. The FoM relative to the $\fnl$ parameter only is shown. The value chosen in the main analysis, $k_{\min} = 10^{-3} h/$Mpc is marked with a triangle. The FoM for the remaining cosmological parameters is largely independent from this choice.
}
\label{fig:kmim}
\end{figure}

We then also look at the effect of changing the large-scale cutoffs $k_{\min}, l_{\min}$. Due to the larger error bars, this is largely unimportant for most cosmological parameters, with the important exception of $\fnl$: due to the form of the scale-dependent bias, the large scales are the region where this parameter is most constrained, and thus the forecasted FoM on $\fnl$ is strongly dependent on this choice, as we can see in Fig.~\ref{fig:kmim} for both the 2D and 3D cases. However, due to the limited size of the surveys, we decide that a conservative choice can hardly push to scales larger than $k_{\min} = 10^{-3} h/$Mpc and $l_{\min} = 5$, which we therefore adopt. A further reason to discard larger scales is to avoid the regime where general relativistic corrections become important, as described e.g. by \citet{Yoo:2010ni}.

\subsection {Effect of the halo model inaccuracies} \label{sec:haloinacc}
\begin {figure} 
\begin{center}
\includegraphics[width=0.4\textwidth,angle=0]{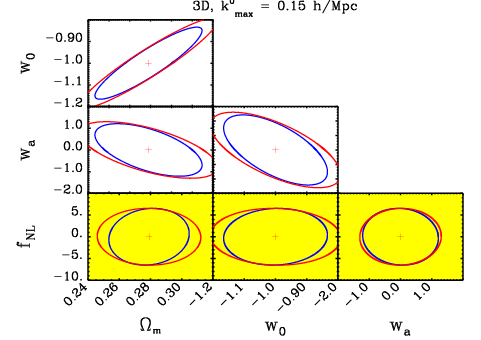}
\vspace{0.2cm}
\includegraphics[width=0.4\textwidth,angle=0]{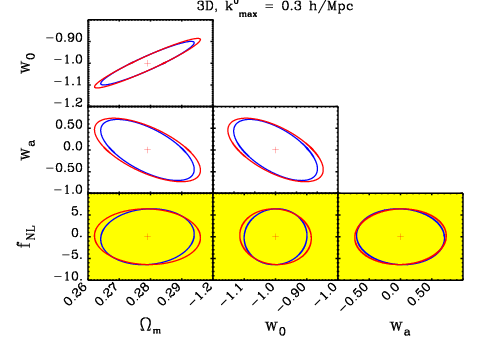}
\end{center}
\caption{Effect of the uncertainties in the modelling of the non-linear regime onto the Fisher matrix forecasts for the 3D power spectrum in the local configuration. The ellipses in blue and red denote here the forecasted posteriors obtained using the halo model and the power spectra from the method by \citet{Smith:2006ne} respectively.  Only scales up to $k_{\max}=0.15$ $h/$Mpc and $k_{\max}=0.3$ $h/$Mpc at $z=0$ were used in the  panels. 
}
\label{fig:testaccu}
\end{figure}

As we have seen in the Section~\ref{sec:nonlin}, the accuracy of the halo model at small scales is only in the range of $10 \%$. For this reason we have tested how severely this impacts on the Fisher matrix forecasts derived using this model. 
For this purpose, we have calculated two sets of Fisher matrices for the Euclid spectroscopic survey, using the 3D power spectrum observable.
For the first matrix of each set, we have derived the power spectra used the halo model; for the second, we have used the approach by \citet{Smith:2006ne}.
 We can see in Fig.~\ref{fig:testaccu} the results using scales up to $k_{\max} = 0.15$ (as used in the main results) and $0.3$ at $z=0$ respectively, from which we can conclude that the 
our results are rather robust in the transition to the non-linear regime.

For weak lensing, where smaller scales are considered, the discrepancy between the two models for the non-linear matter power spectrum grows larger. Fig.~\ref{fig:test_Pk} shows that the model by \citet{Smith:2006ne} deviates from N-body simulations by more than 20\% for $k>2 h/$Mpc, whereas the halo model is significantly more accurate on these scales. Consistently, previous studies have shown that \citet{Smith:2006ne} underestimates the convergence power spectrum by more than 20 \% at $l>1000$, while the halo model is reliable up to $l \sim 50,000$ (see Fig. 9 in \citet{2009A&A...499...31H}). For this reason we only consider the halo model in our weak lensing analysis.

\subsection {Redshift binning}

\begin{figure} 
\begin{center}
\includegraphics[width=0.45\textwidth,angle=0]{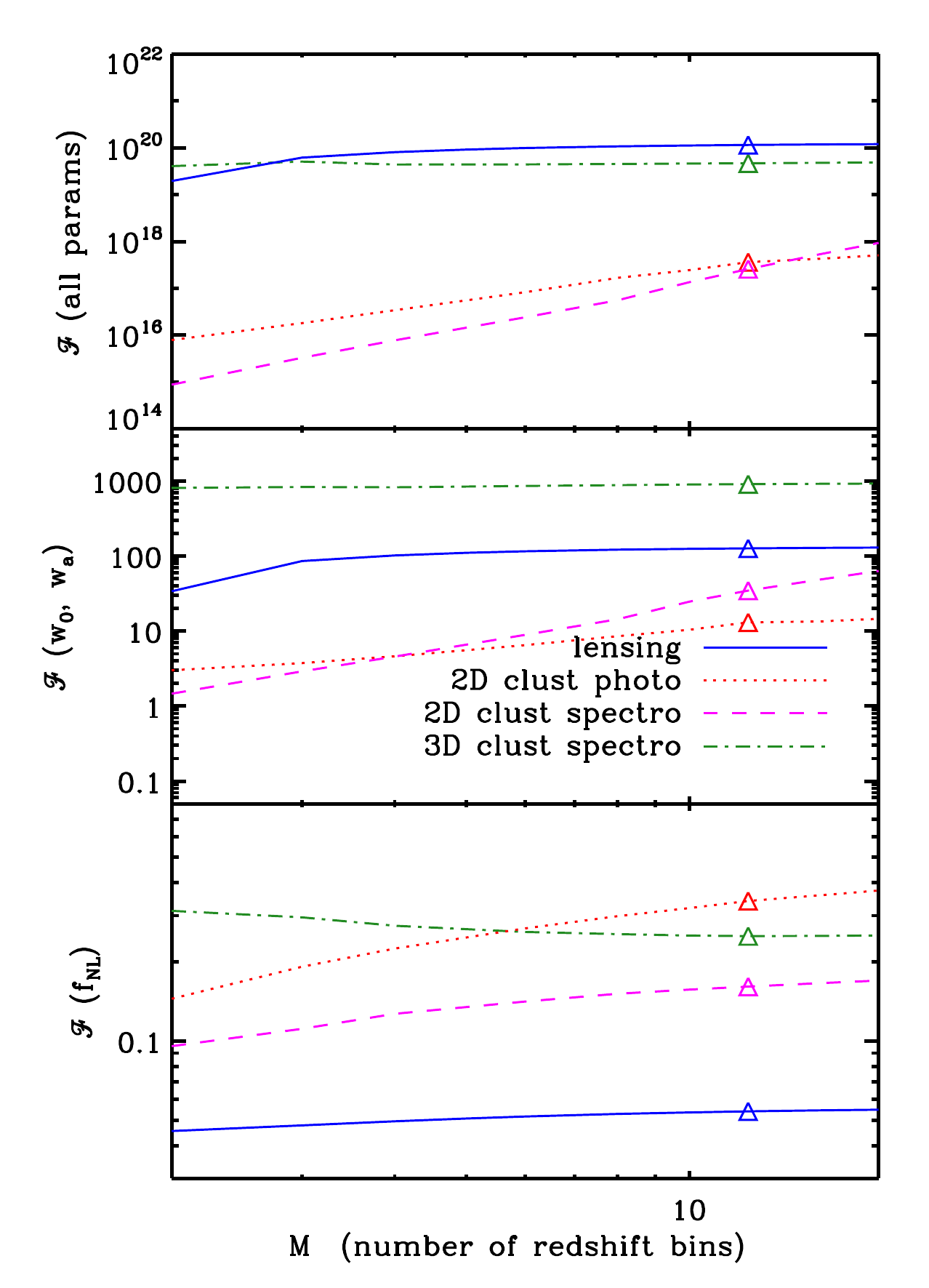}
\end{center}
\caption{Dependence of the Euclid constraints on the redshift tomography. We show the evolution of the figure of merit as a function of the number of redshift bins considered in the analysis, for the all parameters (top), for the dark energy parameters only (middle) and for $\fnl$ (bottom panel). All these results include the Planck priors, and the choice used in the main analysis ($M = 12$) is marked by a triangle. We can see that the results for lensing and the 3D power spectrum have already converged once more than a few bins are used, while for the projected 2D cases the increase in number of bins brings in much additional signal.
}
\label{fig:FoMzbin}
\end{figure}

A further question which may be asked is how important is the choice of the redshift tomography used, i.e. how much would the results change if we used a different binning. We can see in Fig.~\ref{fig:FoMzbin} that this choice is not critical for the lensing and the 3D power spectrum, as in these cases the results have largely converged when we take more than a few bins. On the other hand, for the projected 2D spectra this choice is very important: due to the increasing number of cross-correlations between the bins, the signal-to-noise increases significantly up to a rather high number of bins. For computational reasons, we have decided to carry our analysis using 12 redshift bins, where most cases are quite close to saturation.

\subsection {Euclid Red Book} \label{sec:redbook}

\begin {table*}
\begin {center}
\begin{tabular}{c c c c c c}
\hline
\multicolumn{2}{c}{ Red Book Update }         &   \multicolumn{2}{c}{ Euclid  photometric }  &   \multicolumn{2}{c}{ Euclid spectroscopic }  \\
\hline
Parameter     & description             &   Required & Goal   &   Required & Goal   \\
\hline
$\sigma_z (z)/(1+z)$  & redshift uncertainty  & 0.05  &  0.03      &   0.001    & 0.001     \\
$\bar z$      &median redshift        & 0.8  &  1.0             &  1.0 &  1.1      \\
$n$        &   galaxy density         & 30    arcmin$^{-2}$   &  40    arcmin$^{-2}$         &  1.20 arcmin$^{-2}$    & 1.20   arcmin$^{-2}$       \\
$A$       & surveyed area          & 15 000  sq deg   & 20 000  sq deg         &  15 000  sq deg   & 20 000  sq deg      \\
$dN/dz (z)$ & galaxy distribution         &    \multicolumn{2}{c}{Smail et al.}  &  \multicolumn{2}{c}{Geach et al.  }   \\
\hline
\end{tabular}
\caption{Changes in the specifications for the Euclid Red Book \citep{RedBook}. The parameters which are not shown are unchanged from Table~\ref{tab:techspec}.}
\label{tab:techspec2}
\end{center}
\end {table*}

The detailed specifications of the Euclid mission are still evolving. While the current study was performed based on the so-called Yellow Book \citep{2009arXiv0912.0914L}, the specifications have since evolved to the Red Book \citep{RedBook}. In this section we describe how our results change as a consequence of this update.

We show in Table~\ref{tab:techspec2} the parameters which have changed from the Yellow to the Red book; for the latter, in many cases  two values are given: minimum requirement and goal. As it can be seen, the `goal' specification is generally the same as the Yellow Book value, while the `requirement' is less ambitious. The main exception is the galaxy density in the spectroscopic case, which has increased due to a deeper flux limit.
Table~\ref{tab:RB} reports the changes in the forecasted constraints for the Red book `requirement' specifications for the local PNG case. The Red Book `goal' specifications are very similar to the Yellow book results presented above. We can see that in general the constraints degrade for the photometric survey, and slightly improve for the spectroscopic part, mainly due to the increased number density in this latter case.

In addition, we also show the results obtained using a fiducial bias derived from semianalytic models of galaxy formation \citep{2010MNRAS.405.1006O}, which are available for the spectroscopic case only. In this case the constraints on $\fnl$ degrade further, as this tabulated bias is approximately 15\% lower than the model $b(z) = \sqrt{1+z}$ which we use in our main results.

\begin {table*}
\begin {center}
\begin{tabular}{l c c c c c c c c c}
\hline
  Red Book   & $\sigma (\Omega_{\Lambda})$ & $\sigma (\Omega_b)$ & $\sigma (\Omega_m)$ & $\sigma (h)$ & $\sigma (n)$ & $\sigma (\sigma_8)$ & $\sigma (w_0)$ & $\sigma (w_a)$ & $\sigma (\fnl)$ \\
\hline
lensing &  .078  (.017)  &     .019  (.0019)  &     .016  (.011)  &     .16  (.014)  &     .063  (.0041)  &     .030  (.015)  &     .17  (.12)  &     1.0  (.47)  &     73  (27)  \\
2D clust. phot. &     .095  (.036)  &     .0087  (.0040)  &     .050  (.024)  &     .15  (.030)  &     .043  (.0039)  &     .034  (.026)  &     .50  (.31)  &     1.9  (1.2)  &     5.8  (5.5)  \\
2D clust. spec. &     .076  (.015)  &     .0065  (.0021)  &     .015  (.011)  &     .099  (.015)  &     .050  (.0038)  &     .031  (.015)  &     .090  (.058)  &     .54  (.28)  &     6.4  (5.9)  \\
\textit{---, tab bias} & \textit{.082  (.016)} & \textit{.0069  (.0023)} & \textit{.017  (.012)} & \textit{.11  (.016)} & \textit{.056  (.0039)} & \textit{.036  (.017)} & \textit{.10  (.063)} & \textit{.59  (.30)} & \textit{8.8  (8.0)}  \\
3D + shot noise   &   .067  (.0083)  &     .0048  (.0010)  &     .023  (.0037)  &     .027  (.0064)  &     .033  (.0035)  &     .042  (.017)  &     .13  (.029)  &     .59  (.092)  &     4.1  (4.0)  \\
\textit{---, tab bias} & \textit{.071  (.0084)} & \textit{.0051  (.0010)} & \textit{.025  (.0038)} & \textit{.029  (.0065)} & \textit{.036  (.0035)} & \textit{.048  (.020)} & \textit{.14  (.031)} & \textit{.63  (.098)} & \textit{6.1  (5.8)} \\
\hline
lens. + 2D phot.  &  .030  (.0087)  &     .0056  (.00097)  &     .0048  (.0044)  &     .050  (.0065)  &     .016  (.0036)  &     .0052  (.0048)  &     .054  (.052)  &     .32  (.20)  &     4.7  (4.5)  \\
lens. + 2D spec. &   .030  (.0077)  &     .0052  (.00089)  &     .0051  (.0037)  &     .048  (.0058)  &     .016  (.0036)  &     .0054  (.0045)  &     .051  (.035)  &     .35  (.15)  &     5.7  (5.3)  \\
\textit{---, tab bias} & \textit{.031  (.0080)} & \textit{.0054  (.00092)} & \textit{.0054  (.0040)} & \textit{.051  (.0061)} & \textit{.018  (.0037)} & \textit{.0060  (.0051)} & \textit{.055  (.039)} & \textit{.38  (.16)} & \textit{7.9  (7.3)} \\
\hline
\end{tabular}
\caption{Expected marginalised errors from Euclid for the Red Book specifications (minimum requirements) for the local non-Gaussianity case. The numbers
  within parentheses include the forecasted priors from Planck CMB
  temperature power spectrum. The  3D clustering includes marginalisation over the shot noise parameters in each redshift bin. For the spectroscopic part of the survey, we show both the results obtained using our standard form for fiducial bias $b(z) \propto \sqrt{1+z}$ and the results from the tabulated bias by \citet{2010MNRAS.405.1006O} (in italic). }
\label{tab:RB}
\end{center}
\end {table*}

\section {Conclusions}  \label {sec:concl}

In this work we have studied to what accuracy will the two-point statistics of future surveys of the large-scale structure determine the cosmological parameters, with a particular focus on the non-Gaussianity parameter $\fnl$, considering the most relevant local, equilateral and orthogonal bispectrum configurations.
We have performed a Fisher matrix analysis using the specifications of the upcoming Dark Energy Survey and a Euclid-like survey based on the Euclid Assessment Study Report, using both its spectroscopic and photometric parts. We have chosen these surveys as examples of the future DETF stages III and IV, and the results are likely comparable with other surveys in the same class, such as in particular the planned American mission WFIRST.

We have combined all the relevant data sets, including their covariances. In particular, we have considered the projected 2D galaxy power spectrum, always including curvature and using the exact calculation, discarding the Limber approximation on large scales, as this introduces inaccuracies which are important in the presence of scale-dependent galaxy bias from primordial non-Gaussianity.

We obtained that the strictest constraints on $\fnl$ are expected from the combination of weak lensing and photometric galaxy clustering; in this case we find for the local case $\sigma(\fnl) \simeq 3$ for Euclid and $\simeq 8$ for DES, when also including priors from the temperature power spectra of the Planck CMB mission. In the cases of orthogonal and equilateral configurations, the constraints from galaxy clustering degrade greatly, due to the reduced scale-dependence of the galaxy bias, while the constraints from weak lensing remain at a similar level. Finally, the constraint on scale-dependent non-Gaussianity is for the local case $\sigma(n_{\fnl}) = 0.12$ when the fiducial scale-independent part is $\bar \fnl = 30$.
The level of these constraints is comparable to the expectations from the Planck CMB bispectrum, and fully independent \citep{Komatsu:2001rj,Sefusatti:2009xu}.

We have also studied the effect of updating the description of Euclid to the latest Red Book specifications; in this case using the Red Book `goal' parameters leaves the forecasts largely unchanged, while we found that using the `requirement' parameters degrades the constraints on PNG to $\sigma (\fnl) \simeq 5 $.

Further extensions of this approach, which will be useful not only for forecasting but also for the likelihood analysis of the upcoming real data, will include on one hand expanding the total covariance of the two-point statistics to include observations from clusters of galaxies and the correlations with the cosmic microwave background; on the other hand, this approach will be extended to the inclusion of the three- and four-point statistics, which will be instrumental in the search for the higher-order $g_{\mathrm{NL}} $ and $ \tau_{\mathrm{NL}}$ inflationary parameters.

\section*{Acknowledgements}

We acknowledge Jochen Weller for supplying us with the CMB Fisher
matrix code, and Christian Byrnes, Martin Kilbinger, Fabian Schmidt and Masahiro Takada for useful discussions.  We thank the Euclid collaboration, Yannick Mellier and in particular the
theory working group for stimulating discussions.
TG acknowledges support from the Alexander von Humboldt Foundation.
JC and AP acknowledge support from the Swiss Science National Foundation.
CP acknowledges support from the German Research Association (DFG)
through the Transregional Collaborative Research Center ``The Dark Universe''
(TRR33).

\appendix

\section {Primordial non-Gaussianity and the large-scale structure} \label{sec:appbias}

We review here the effects of primordial non-Gaussianity on the large-scale structure, focussing on the results which are used in this analysis.

\subsection {Bispectra and skewness}

The bispectrum of the Bardeen potential $\Phi$ in the local, equilateral and orthogonal cases is given by \citep{Taruya:2008pg,Schmidt:2010gw}
\be \label{eq:bloc}
B^{\mathrm{loc}}_{\Phi} (k_1, k_2, k_3) \simeq 2 \fnl \left[ P_{\Phi}(k_1) P_{\Phi}(k_2) + 2 \, \mathrm{perms.}\right] \, ,
\ee
\ba \label{eq:beq}
B^{\mathrm{equ}}_{\Phi} (k_1, k_2, k_3) & \simeq & 6 \fnl^{\mathrm{eq}} \, \Bigl{\{}  - \left[ P_{\Phi}(k_1) P_{\Phi}(k_2) + 2 \, \mathrm {perms.} \right] \Bigr. \nonumber \\
&~& \left.  - 2 \left[P_{\Phi}(k_1) P_{\Phi}(k_2) P_{\Phi}(k_3) \right]^{2/3} \right.  \nonumber \\
&~& + \left.  \left[ P_{\Phi}^{1/3}(k_1) P^{2/3}_{\Phi}(k_2) P_{\Phi}(k_3)  + 5 \, \mathrm{perms.}\right] \, \right\} \, ,
\ea
\ba \label{eq:borth}
B^{\mathrm{ort}}_{\Phi} (k_1, k_2, k_3) & \simeq & 6 \fnl^{\mathrm{ort}} \, \Bigl{\{}  -3 \left[ P_{\Phi}(k_1) P_{\Phi}(k_2) + 2 \, \mathrm{perms.} \right] \Bigr. \nonumber \\
&~& \left. - 8 \left[P_{\Phi}(k_1) P_{\Phi}(k_2) P_{\Phi}(k_3) \right]^{2/3} \right.  \nonumber \\
&~& + \left. 3 \, \left[ P_{\Phi}^{1/3}(k_1) P^{2/3}_{\Phi}(k_2) P_{\Phi}(k_3)  + 5 \, \mathrm{perms.}\right] \, \right\} \, .
\ea
It is interesting to compute the lowest moments of
the corresponding linear density field $\delta$: the variance and the
skewness. 
The variance is defined as
\be
\sigma^2 (M) \equiv \langle \delta_M^2 \rangle = \frac{1}{2 \pi^2} \int dk \, k^2 P(k) F^2(k,M) \, ,
\ee
where we have introduced the smoothed field $\delta_M(k) = \delta(k) F (k,M)$, and $F(k,M)$ is a filter function of mass resolution $M$. We use a top-hat function in real space.

The third momentum of the smoothed density field can be written in terms of the three-point function
\be
\langle \delta^3_M \rangle = \int \frac {d^3 k_1}{(2 \pi)^3} \int \frac {d^3 k_2}{(2 \pi)^3} \int \frac {d^3 k_3}{(2 \pi)^3} \langle  \delta_M (\mathbf{k}_1) \delta_M (\mathbf{k}_2) \delta_M (\mathbf{k}_3)  \rangle \, .
\ee
By substituting the bispectra from Eqs.~(\ref{eq:bloc}, \ref{eq:beq}, \ref{eq:borth}), and integrating the Dirac delta to give $\mathbf{k}_3 = -\mathbf{k}_1 - \mathbf{k}_2$ so that we can define $k \equiv  | \mathbf{k}_1 + \mathbf{k}_2| $,
 this expression can be simplified for the three cases extending the calculation by \citet{2009MNRAS.396...85D}
\ba
\langle \delta^3_M \rangle^{\mathrm{loc}} &=& \frac {\fnl} {(2 \pi^2)^2} \int_0^{\infty} d k_1 k_1^2 \frac {P (k_1)} {\alpha (k_1)} \, \int_0^{\infty} d k_2 k_2^2 \frac {P (k_2)} {\alpha (k_2)} \nonumber \\
&~& \times \int_{-1}^1 d \mu  \alpha (k) \left[ 1 + 2 \frac {P(k) \alpha^2(k_2)}{P(k_2) \alpha^2(k)}  \right] \, \nn \\
&~& \times F(k_1,M)\,F(k_2,M)\,F(k,M)\,  ,
\ea
\ba \label{eq:skewequi}
\langle \delta^3_M \rangle^{\mathrm{equ}} &=& \frac {3 \fnl} {(2 \pi^2)^2} \int_0^{\infty} d k_1 k_1^2  \alpha (k_1) \, \int_0^{\infty} d k_2 k_2^2  \alpha (k_2) \,  \nonumber \\
&~& \times \, \int_{-1}^1 d \mu  \, \alpha (k)    \left\{  - \left[ \frac {P(k_1) P(k_2)} {\alpha^2(k_1) \alpha^2(k_2)} + 2 \, \mathrm{perms.} \right]  \phantom{\sum} \right. \nonumber \\
&~&  \left.  - 2 \left[ \frac {P(k_1) P(k_2) P(k)} {\alpha^2(k_1) \alpha^2(k_2) \alpha^2(k)} \right]^{2/3} \right. \nonumber \\
&~&  \left.  + \left[ \frac {P^{1/3}(k_1) P^{2/3}(k_2) P(k)} {\alpha^{2/3}(k_1) \alpha^{4/3}(k_2) \alpha^2(k)}  + 5 \, \mathrm{perms.}\right] \, \right\}\, \, \nn \\
&~& \times F(k_1,M)\,F(k_2,M)\,F(k,M)\,  ,
\ea
\ba \label{eq:skeworth}
\langle \delta^3_M \rangle^{\mathrm{ort}} &=& \frac {3 \fnl} {(2 \pi^2)^2} \int_0^{\infty} d k_1 k_1^2  \alpha (k_1) \, \int_0^{\infty} d k_2 k_2^2  \alpha (k_2) \,  \nonumber \\
&~& \times \, \int_{-1}^1 d \mu  \, \alpha (k)    \left\{  -3 \left[ \frac {P(k_1) P(k_2)} {\alpha^2(k_1) \alpha^2(k_2)} + 2 \, \mathrm{perms.} \right]  \phantom{\sum} \right. \nonumber \\
&~&  \left.  - 8\left[ \frac {P(k_1) P(k_2) P(k)} {\alpha^2(k_1) \alpha^2(k_2) \alpha^2(k)} \right]^{2/3} \right. \nonumber \\
&~&  \left.  + 3 \left[ \frac {P^{1/3}(k_1) P^{2/3}(k_2) P(k)} {\alpha^{2/3}(k_1) \alpha^{4/3}(k_2) \alpha^2(k)}  + 5 \, \mathrm{perms.}\right] \, \right\}\, \, \nn \\
&~& \times F(k_1,M)\,F(k_2,M)\,F(k,M)\, ,
\ea
The skewness is finally defined for all cases as
\be
S_3(M) \equiv \langle \delta^3_M \rangle / \langle \delta^2_M \rangle^2 = \langle \delta^3_M \rangle / \sigma^4 \, .
\ee

\subsection {Mass function}
The halo mass function $n(M,z)$ gives
the number density of haloes of mass $M$ at redshift $z$.
Various analytical methods have been developed to compute this quantity
starting from the statistical properties of Gaussian primordial perturbations, starting with \citet{Press:1973iz}. 
Fitting formulae that improve agreement with N-body simulations have also
been presented \citep{Sheth:1999mn, Sheth:2001dp, Jenkins:2000bv, Warren:2005ey, Tinker:2008ff, Pillepich:2008ka}.
 
The halo mass function is modified by the presence of PNG
\citep{Matarrese:2000iz, LoVerde:2007ri, Maggiore:2009rx, 2009MNRAS.398.2143L}.
Since analytical models are based on a set of simplistic assumptions,
they most robustly predict the ratio between
the mass function generated by non-Gaussian and Gaussian initial conditions
with the same power spectrum
(see \citet{Giannantonio:2009ak} for a comparison of the different methods
against N-body simulations). 

For this reason, in our
forecasts we use the fitting formula given by PPH08 for the Gaussian mass
function and multiply it by the factor \citep{LoVerde:2007ri} 
\ba
R_\mathrm{LV}  \left( \frac {\delta_c} {\sigma}, \fnl \right) &\equiv&
 \left[ 1 + \frac {S_3  \sigma^2} {6 \sqrt q \, \delta_c}  \left( \frac{q^2 \, \delta_c^4} {\sigma^4} -   \frac{2 q\, \delta_c^2} { \sigma^2 } - 1 \right) + \right. \nonumber \\
&~& + \left. \frac {dS_3(\sigma)}{d \mathrm{ln} \sigma^2} \frac{\sigma} {6 \sqrt q \, \delta_c}    \left( \frac{q\, \delta_c^2}{\sigma^2} - 1 \right) \right] \, , 
\label{LVratio} 
\ea
in the non-Gaussian case.
Here $\delta_c\simeq 1.686$ is the threshold for the collapse of a linear
density perturbation and the
corrective factor $q$, of order unity, is a heuristical correction which may be applied to the collapse threshold to improve
the agreement with N-body simulations, and which we set to unity for the reasons given in Section \ref{sec:NG}. Note that PNG enters the expression for the
halo mass function through the linear density skewness $S_3(M)$ which
is linearly proportional to $\fnl$.

\subsection {Halo bias}

The halo linear bias factors may be obtained from the expression for the halo mass function using the peak-background split formalism \citep {Bardeen:1985tr, Cole:1989vx, Mo:1995cs, Catelan:1997qw}. In the Gaussian case, the bias is obtained by taking the logarithmic derivative of the mass function with respect to the
collapse threshold. This gives a mass and redshift  dependent bias coefficient.

Recent generalisations of the peak-background split to cases with PNG 
have shown that the corrections to the mass function generate two 
extra terms in the expression for the halo bias: one of them is a small 
addition to the Gaussian bias while the second one introduces a 
scale-dependent term
(see Eqs. (\ref{biasPNG}) and (\ref{eq:dball}) in the main text)  
\citep{Dalal:2007cu,Slosar:2008hx,Matarrese:2008nc,Afshordi:2008ru,Giannantonio:2009ak, Schmidt:2010gw,2011PhRvD..84f1301D}.

\section {Planck Fisher matrix} \label{app:planck}

The Planck Fisher matrix was kindly made available to us by Jochen Weller. It was calculated according to \citet{Albrecht:2009ct, Rassat:2008ja} and used in \citet{Sartoris:2010cr}. As described in Section 3.8 of \citet{Sartoris:2010cr}, it is computed adopting the parameterisation $\mathbf{\vartheta} = (\omega_m, \vartheta_s , \ln A_S , \omega_b, n_s , \tau)$: with $A_s$ the amplitude of primordial perturbations,
$\vartheta_s$ the size of the sound horizon at the last-scattering surface,  and $\tau$ the optical depth to Compton scattering which has
been marginalised over.
The calculation of the Fisher matrix is based on the conservative assumption that only the 143 Ghz channel is used for science analysis. In this case, the beam size is $\vartheta_{\mathrm{FWHM}} = 7.1' $, and the temperature and polarisation sensitivities are  $\sigma_T = 2.2 \mu \mathrm{K}/\mathrm{K}$ and $\sigma_P = 4.2 \mu \mathrm{K}/ \mathrm{K}$, respectively. In order to avoid polarisation foregrounds, only multipoles with $l_{\min} = 30$ have been considered together with a sky fraction  $f_\mathrm{sky} = 0.8$ to minimise the effects of galactic foregrounds.
The transformation of the Fisher matrix for the parameter set $(\Omega_m, \Omega_{\Lambda}, h, \sigma_8, \Omega_b,w_0,w_a, n_s )$
is done as in \citet{Rassat:2008ja}.
It is important to highlight that the Planck Fisher matrix does not include any information on primordial non-Gaussianity, as CMB constraints on PNG can be calculated from higher-order statistics only (especially the bispectrum). So adding these priors will not improve the constraints on $\fnl$.

 \bibliographystyle{mn2e}
 \bibliography{ms}

\label{lastpage}

\end{document}